\documentclass[a4paper,12pt]{article}

\pdfoutput=1

\usepackage{jheppub}
\usepackage[table,dvipsnames]{xcolor}
\usepackage{graphicx}
\usepackage{float}         
\usepackage{bm}        
\usepackage{amsmath}
\usepackage{etoolbox}
\usepackage{enumitem}
\usepackage{stackrel}
\usepackage[normalem]{ulem}
\usepackage{mathtools}
\usepackage{nicefrac}
\usepackage{units}
\usepackage[T1]{fontenc}                            
\usepackage[utf8]{inputenc}
\usepackage{lmodern,mathrsfs}
\usepackage{amssymb,amsthm,amsfonts}   
\usepackage{physics}
\usepackage{titlesec}
\usepackage{csquotes}
\usepackage{listings}

\makeatletter
\patchcmd{\maketitle}{\@fpheader}{}{}{}
\makeatother
\newcommand{\mathsym}[1]{{}}

\usepackage{amsmath,amssymb}

\usepackage{bm} 

\newsavebox{\notrightarrow}
\sbox{\notrightarrow}{$\to$\hspace{-4mm}/}

\newsavebox{\PARTIALSLASH}
\sbox{\PARTIALSLASH}{$\partial$\hspace{-1.6mm}/}

\newsavebox{\ASLASH}
\sbox{\ASLASH}{$A$\hspace{-2.1mm}/}

\newsavebox{\KSLASH}
\sbox{\KSLASH}{$k$\hspace{-1.8mm}/}

\newsavebox{\LSLASH}
\sbox{\LSLASH}{$\ell$\hspace{-1.8mm}/}

\newsavebox{\QSLASH}
\sbox{\QSLASH}{$q$\hspace{-1.8mm}/}

\newsavebox{\DSLASH}
\sbox{\DSLASH}{$D$\hspace{-2.2mm}/}

\newsavebox{\DbfSLASH}
\sbox{\DbfSLASH}{${\mathbf D}$\hspace{-2.8mm}/}

\newsavebox{\DELVECRIGHT}
\sbox{\DELVECRIGHT}{$\stackrel{\rightarrow}{\partial}$}

\newcommand{\blue}{\IfColor{\textCadetBlue}{}}
\newcommand{\black}{\IfColor{\textBlack}{}}
\newcommand{\red}{\IfColor{\textRed}{}}
\newcommand{\green}{\IfColor{\textOliveGreen}{}}
\newcommand{\lila}{\IfColor{\textRedViolet}{}}

\AtBeginEnvironment{quote}{\itshape}

\titleformat
{\part} 
[display] 
{\bfseries\Huge} 
{\itshape Chapter.~\thepart} 
{0.5ex} 
{
	\rule{\textwidth}{1pt}
	\vspace{1ex}
	\centering
} 
[
\vspace{-0.5ex}%
\rule{\textwidth}{0.3pt}
] 
\usepackage{mathbbol}
\usepackage{color}
\usepackage{natbib}
\usepackage[english]{babel}
\usepackage{ragged2e}
\newcommand{\diag}{\mathop{\rm diag}\nolimits}
\newcommand{\rd}[1]{\mathop{\mathrm{d}#1}}

\newcommand{\pderiv}[2]{\frac{\partial#1}{\partial#2}}

\newcommand{\intx}[2]{\int \mathrm{d}^#2#1\,}

\newcommand{\inv}[1]{\frac{1}{#1}}

\newcommand{\tder}[1]{\mathfrak{D}#1}

\newcommand{\dotpr}[2]{{#1}\cdot{#2}}
\newcommand{\cdbot}[2]{\nabla^\bot_{#1}{#2}}

\newcommand{\dual}[1]{{^\star{#1}}}

\DeclareBoldMathCommand{\xbot}{{\bf x}_\perp}
\newcommand{\tvec}[1]{\boldsymbol{#1}}

\DeclareRobustCommand{\ed}{\ensuremath{e}}

\newcommand{\half}{\frac{1}{2}}

\newcommand{\mcA}{{\mathcal{A}}}
\newcommand{\mcB}{{\mathcal{B}}}

\newcommand{\mcJ}{{\mathcal{J}}}

\newcommand{\mcL}{{\mathcal{L}}}

\newcommand{\mcR}{{\mathcal{R}}}

\newcommand{\rfl}[1]{Equation~(\ref{#1})}

\newcommand{\rfsec}[1]{Sec.~\ref{#1}}

\newcommand{\rff}[1]{Fig.~\ref{#1}}

\newcommand{\rfApp}[1]{Appendix~\ref{#1}}

\newcommand{\nn}{\nonumber}
\def\a{\alpha}
\def\b{\beta}

\newcommand{\p}{\partial}

\newcommand{\f}[2]{\frac{#1}{#2}}

\newcolumntype{L}[1]{>{\raggedright\let\newline\\\arraybackslash\hspace{0pt}}m{#1}}
\newcolumntype{C}[1]{>{\centering\let\newline\\\arraybackslash\hspace{0pt}}m{#1}}
\newcolumntype{R}[1]{>{\raggedleft\let\newline\\\arraybackslash\hspace{0pt}}m{#1}}

\definecolor{lightgray}{gray}{0.9}
\preto\tabular{\setcounter{magicrownumbers}{0}}
\newcounter{magicrownumbers}

\newcommand{\bea}{\begin{eqnarray}}
	\newcommand{\beal}[1]{\begin{eqnarray}\label{#1}}
		\newcommand{\eea}{\end{eqnarray}} 
	\newcommand{\be}{\begin{equation}} 
		\newcommand{\bel}[1]{\begin{equation}\label{#1}}
			\newcommand{\ee}{\end{equation}}

		\newcommand{\xv}{{\boldsymbol x}} 
		
		\def\epsLmnab{\epsilon_{\mu\nu\alpha\beta}}
		\def\epsUmnab{\epsilon^{\mu\nu\alpha\beta}}

		\def\S0iU{{\Sigma}^{0i}}

		\def\n0{n_{0}}
		\def\e0{\varepsilon_{0}}
		\def\P0{P_{0}}
		
		\def\ed{\varepsilon}	         

		\renewcommand\sout{\bgroup\markoverwith{\textcolor{blue}{\rule[0.5ex]{2pt}{0.4pt}}}\ULon}

		\newcommand{\axion}{\mathrm{\Theta}}
		
		\def\eMf{electromagnetic (EM)}
		\def\EM{EM}

		\def\LRF{LRF}

		\def\qGPf{quark gluon plasma (QGP)}
		\def\QGP{QGP}

		\def\MHD{MHD}
		\def\iMHD{iMHD}
		
		\def\hD{hydrodynamics}
		\def\RHD{Relativistic hydrodynamics}
		\def\rHD{relativistic hydrodynamics}
		\def\rMHDf{relativistic magnetohydrodynamic (RMHD)}
		
		\def\RMHD{RMHD}
		
		\def\CS{CS}
		\def\MCSf{Maxwell-Chern-Simons (MCS)}
		\def\MCS{MCS}
		\def\CSMHDf{Chern-Simons magnetohydrodynamics (CSMHD)}
		\def\CSMHD{CSMHD}
		\def\cMEf{Chiral magnetic effect (CME)}
		
		\def\CME{CME}
		\def\cMf{chiral magnetic (CM)}
		
		\def\CM{CM}
		
		\def\aHf{anomalous Hall (AH)}
		
		\def\AH{AH}
		\def\Alfven{Alfv\'en}

		\def\mcs{\ensuremath{\text{MCS}}}
		\def\maxwell{\ensuremath{\text{Maxwell}}}

		\def\EOM{EOM}
		\def\fl{\ensuremath{\text{Fluid}}}
		\def\lrf{\ensuremath{\text{LRF}}}
		\newcommand{\CA}{\ensuremath{C_A}}
		\newcommand{\vs}{\ensuremath{v_s}}
		\newcommand{\val}{\ensuremath{v_a}}
		\newcommand{\DT}{\ensuremath{\delta T}}
		\newcommand{\PROJ}{\ensuremath{\Delta}}
		\newcommand{\PR}{\ensuremath{p}}
		\newcommand{\DP}{\ensuremath{\delta p}}
		\newcommand{\EP}{\ensuremath{\epsilon}}
		\newcommand{\DEP}{\ensuremath{\delta\epsilon}}

		\newcommand{\DAX}{\ensuremath{\delta\axion}}
		
		\newcommand{\MUF}{\mu_5}

		\newcommand{\JT}{{\mcJ}}
		\newcommand{\PREQ}{\ensuremath{p_{0}}}
		\newcommand{\EPEQ}{\ensuremath{\epsilon_{0}}}

		\newcommand{\ELMU}{\ensuremath{\mu_{e}}}
		\newcommand{\ELMUEQ}{\ensuremath{\mu_{e,0}}}
		\newcommand{\ELCH}{\ensuremath{n_{e}}}
		\newcommand{\ELCHEQ}{\ensuremath{n_{e,0}}}
		\newcommand{\ELCOND}{\ensuremath{\sigma_{e}}}
		\newcommand{\ELCONDEFF}{\ensuremath{\sigma_{\rm eff.}}}
		\newcommand{\FUG}{\ensuremath{\alpha}}
		
		\newcommand{\RAXION}{\ensuremath{\mcR_{\axion}}}

		\def\epsLmnab{\epsilon_{\mu\nu\alpha\beta}}
		\def\epsUmnab{\epsilon^{\mu\nu\alpha\beta}}

		\def\S0iU{{\Sigma}^{0i}}

		\def\n0{n_{0}}
		\def\e0{\varepsilon_{0}}
		\def\P0{P_{0}}
		
		\def\ed{\varepsilon}	         

		\newcommand{\polar}{\theta}
		\newcommand{\azim}{\phi}
		\newcommand{\uvx}{\tvec{\hat{x}}}
		\newcommand{\uvy}{\tvec{\hat{y}}}
		\newcommand{\uvz}{\tvec{\hat{z}}}
		\newcommand{\uvkbot}{\tvec{\hat{k}_\bot}}
		\newcommand{\uvkpar}{\tvec{\hat{k}_\parallel}}
		\newcommand{\ncch}{\ensuremath{\mathcal{D}_{\text{NC}}}}
		\newcommand{\cch}{\ensuremath{\mathcal{D}_{\text{C}}}}
		\def\vg{v_g}
\begin{document}
\title{
	Anomalous Hall instability in the Chern-Simons magnetohydrodynamics
}
\author{M. Kiamari,}
\author[a]{M. Rahbardar,}
\author[b]{M. Shokri,}
\author[c]{and N. Sadooghi}
\emailAdd{m\_kiamari@ut.ac.ir, mehrnoosh.rahbardar@student.sharif.edu,  mshokri@ipm.ir (Corresponding author), sadooghi@physics.sharif.ir}
\affiliation[a,c]{Department of Physics, Sharif University of Technology, 
	P.O. Box 11155-9161, Tehran, Iran}
\affiliation[b]{School of Particles and Accelerators, IPM,  P.O. Box 19395-5531, Tehran, Iran}
\abstract{
The \CSMHDf\,is introduced using a \MCSf\,Lagrangian including an axion-like field $\Theta$. The MCS equation of motion derived from this Lagrangian consists of a modified current, including a \cMf\,and an \aHf\,current, in addition to the ordinary Ohm current of resistive magnetohydrodynamics (MHD). The former consists of an axial chemical potential, which is given in terms of the temporal comoving derivative of $\Theta$, and the latter arises from the spatial gradient of $\Theta$. As it turns out, the existence of {the axial chemical} potential is a nonequilibrium effect that plays no role in the linear stability analysis, whereas the \AH\,current arises as in the first-order linear perturbation of the thermal equilibrium. We analyze the linear stability and causality of the \CSMHD\,in a resistive and chiral medium. We show that the \Alfven\,modes propagating sufficiently close to the direction of the magnetic field are unstable but causal. They are also accompanied by a genuine nonhydro mode.  A stable mode in a particular direction can correspond to an unstable mode propagating in the exact opposite direction. The AH instability is a manifestation of a breakdown of the parity. {A numerical analysis of the phase velocity confirms these results}.
}
\arxivnumber{}
\pagestyle{myplain}

\maketitle
\toccontinuoustrue
\flushbottom

\section{Introduction}\label{sec:intro}
\setcounter{equation}{0}
The successes of \rHD\,\cite{Landau1987Fluid,rezzolla} in explaining the experimental data from the heavy-ion collisions\,\cite{Schenke2010,Schenke2011,Paquet:2015lta} have led to overgrowing theoretical attention toward it\,\cite{Florkowski:2017olj}. In particular, the stability and causality of dissipative \rHD\,\cite{Hiscock:1983zz,Hiscock:1985zz,Kovtun:2019hdm}, is crucial in numerical simulations. The \textit{unreasonable effectiveness} of stable and causal formulations of dissipative \rHD\,outside the equilibrium\,\cite{Romatschke:2016hle}, motivated the discovery of \hD\,attractors\,\cite{Heller:2015dha}. \RHD\,is a universal theory in the sense that the underlying microscopic nature of the system appears only through different transport coefficients\,\cite{Kovtun:2012rj}. The same remark holds for MHD, which couples the dynamics of a conducting fluid with {the} Maxwell equations\,\cite{Bekenstein1978}. For the fluids in which the magnetic Reynolds number is small\,\cite{Huang:2015oca}, a resistive formulation of \rMHDf\,is well motivated. Resistive \RMHD\,can also be formulated on universal grounds\,\cite{Denicol:2019iyh}. However, a macroscopic description of certain phenomena with quantum nature cannot be incorporated into the ordinary formulation of \rHD\,through a mere modification of transport coefficients. In particular, anomalous transport in the chiral matter and macroscopic effects of spin are so\,\cite{Newman:2005hd,Son2009,Florkowski:2017ruc,Sadooghi:2016ljd}.
\par
In {a} chiral {fluid}  the number of left- and right-handed fermions is {locally} unequal. In the \qGPf, such chiral imbalance is a result of the interplay of the chiral anomaly and nontrivial gluon configurations\,\cite{Kharzeev:2007jp,Fukushima:2008xe}.
To {describe} {this} fluid {using} the chiral \MHD\,theory, {one} introduces new terms into the current~\cite{Newman:2005hd,Hattori:2017usa}. The new terms may at least include a \CM\,vector current and an axial current. These modifications of the current give rise to new terms in the entropy current\,\cite{Son2009}. The second law of thermodynamics is then employed to find constraints on the new transport coefficients, {including the anomalous ones \cite{Son2009, Sadooghi:2016ljd}.} The {non-relativistic and relativistic} chiral \MHD\,is vastly investigated in the literature\,\cite{Boyarsky2015,Rogachevskii:2017uyc,Hattori:2017usa,Shovkovy:2018tks,Sadooghi2018}. In particular, the existence of unstable propagating modes has been studied {in \cite{Rogachevskii:2017uyc,Boyarsky2015,Hattori:2017usa}.}
\par
In the present work, we take another approach based on the \MCS\,Lagrangian. The \MCS\,Lagrangian adds a topological term, {including} an axion-like field, to the ordinary Maxwell theory\,\cite{Kharzeev:2009fn}. The axion-like field connects the \eMf\,fields to the topological properties of the matter, and gives rise to the equation of axion electrodynamics. Novel phenomena such as the \cMEf\,\cite{Kharzeev:2007jp,Fukushima:2008xe}, the Witten effect\,\cite{Witten:1979ey}, and photon's topological mass naturally arise from the axion electrodynamics\,\cite{Wilczek1987,Carroll:1989vb,Kharzeev:2009fn,Landsteiner:2016led}. One can also extract an extension of the \MHD\,from \MCS\,theory that includes anomalous transport without an ad-hoc modification of the current\,\cite{Ozonder:2010zy,Ozonder-Erratum,Sadooghi2018}. As a result, and in contrast to the chiral \MHD, there is no axial current in {this so-called}  \CSMHD. It has been shown that the second law of thermodynamics in \MCS\,theory is a consequence of the modification of thermodynamic relations in {the} presence of the axion-like field\,\cite{Ozonder:2010zy,Ozonder-Erratum}. Following this observation, we derive the entropy current, and define the equilibrium state of {an}  electrically neutral fluid. We then perturb the equilibrium to investigate the stability and causality of \CSMHD. Although in the resistive \MHD, the electric conductivity may not be large enough to suppress the electric field and ensure neutrality, we assume the electric field and vector chemical potential vanish in equilibrium. Such a power-counting scheme is {referred to as}  the weak electric field regime {(see  \cite{Kovtun:2016lfw,Hernandez:2017mch})}. The axial {chemical} potential, {defined by the time derivative of the axion-like field,} is a non-equilibrium effect, and does not appear in the linear stability treatment. However, the spatial inhomogeneity of the axion-like field at a macroscopic level leads to a novel \AH\,effect at the first order of fluctuations {\cite{Kharzeev:2009fn, Sadooghi2018}}. The combination of the \AH\,and Ohm currents are effectively understood through the definition of an effective conductivity, {which, as it turns out,} does not have a definite sign. {We show that its} sign and value depend on the direction that a mode propagates. The indefiniteness of the effective conductivity sign gives rise to unstable but causal \Alfven\,modes\,{originally introduced in \cite{Gedalin1993}}, propagating close to the plane transverse to the magnetic field. {We also show that the} \Alfven\,modes are also manifestly affected by the breakdown of rotational and parity symmetries. In addition to the modification of magnetosonic and \Alfven\,waves, we find two genuine nonhydro (gapped) modes. In contrast to the \Alfven\,modes, the magnetosonic and their accompanying nonhydro mode are linearly stable for the full range of wavenumbers, and are not affected by the chirality. Unsurprisingly, the magnetosonic modes are damped by resistivity. Although the emergence of the unstable mode has been observed in chiral MHD\,\cite{Hattori:2017usa}, the source and nature of \AH\,instability {discussed in the present paper} are different: {Whereas} in chiral \MHD, the instability is induced by the axial charge, in \CSMHD, it is a result of the change of topological charges in neighboring domains. We also do not employ any approximating constraint on the values of wavenumbers and electrical resistivity.
\par
The organization of this paper is as follows. In \rfsec{sec:remarks}, we review the equations of motion in the MCS theory and its thermodynamics. Then, we fix the hydrostatic configuration. In \rfsec{sec:modes-lrf}, we implement linear treatment to investigate the stability and causality of the theory. In \rfsec{sec:numerics}, we present a numerical investigation of the phase velocities and the imaginary parts of the different modes,
 The paper is concluded in \rfsec{sec:conc}. We use the natural units in which $\hbar=c=k=1$. The convention of the metric signature is mostly minus, namely $g_{\mu\nu}=\diag\left(+1,-1,-1,-1\right)$. A review of some useful relations is presented in \rfApp{app:notation}.
\section{General remarks}\label{sec:remarks}
\subsection{ {Review material}}
\setcounter{equation}{0}
\par
In this section, we review the equations of the \CSMHD\,theory {and derive the stationary solution to its \EOM\,in thermal equilibrium. We refer to this solution as the hydrostatic configuration of \CSMHD.} This theory is based on the \MCS\,Lagrangian density \cite{Kharzeev:2009fn}
\begin{equation}\label{eq:mcs-lagrangian}
	\mcL_{\mcs} =\mcL_{\maxwell}+\mcL_{\text{CS}},
\end{equation}
with
\begin{eqnarray*}
\mcL_{\maxwell}\equiv -\inv{4}F^{\mu\nu}F_{\mu\nu} - A_\mu J^\mu,\qquad \mbox{and}\qquad
\mcL_{\text{CS}}\equiv - \frac{\CA}{4}\Theta F^{\mu\nu}\dual{F}_{\mu\nu}.
\end{eqnarray*}
{In $\mcL_{\text{CS}}$, $\Theta(x)$ is the axion-like field, and the field strength tensor $F_{\mu\nu}$ and its dual $\dual{F}_{\mu\nu}$ are given by
\begin{equation}
	F_{\mu\nu} = \p_\mu A_\nu - \p_\nu A_\mu,\qquad \mbox{and}\qquad \dual{F}_{\mu\nu} = \half \epsilon_{\mu\nu\alpha\beta} F^{\alpha\beta}.
\end{equation}}
The anomaly coefficient $\CA$ before the topological term $F^{\mu\nu}\dual{F}_{\mu\nu}$ in \eqref{eq:mcs-lagrangian} reads
\begin{equation}
	\CA = N_C \sum_{\text{f}}q_f^2\frac{e^2}{2\pi^2},
\end{equation}
with $\sum_{\text{f}}$ being the summation over the quark flavors {with charge $q_f$} and $N_C$ the number of colors. {The number of the quark flavors depend on the energy scale of the system under consideration}. In \eqref{eq:mcs-lagrangian}, $J^\mu$ {is the ordinary electromagnetic (\EM\,) current, and is determined by taking the functional derivative of $\mcL_{\maxwell}$ with respect to the \EM\,source $A_{\mu}$.} {Taking, however, the variation of the full \MCS\,action
$\mcL_{\mcs}$ with respect to $A_{\mu}$, an additional term proportional to $C_{A}$ appears in the resulting current $\JT^{\mu}$ ,}
\begin{equation}\label{eq:mcs-current}
	\JT^{\mu} = J^\mu + \CA\dual{F}^{\mu\nu}P_{\nu},  	
\end{equation}
{where $P_\mu \equiv \p_\mu \axion$. }
The emergence of {$\JT^{\mu}$} is a consequence of {the} spacetime dependency of $\axion$, and gives rise to a modification of {the homogeneous and inhomogeneous} Maxwell equations,
\begin{eqnarray}\label{eq:mcs}
	\p_\mu \dual{F}^{\mu\nu} = 0,\qq{and} \p_\mu F^{\mu\nu} = \JT^\nu,
\end{eqnarray}
{as well as the energy-momentum conservation relation,}
\bel{eq:csmhd}
\p_\mu T_{\fl}^{\mu\nu} = F^{\nu\lambda}J_\lambda - \frac{\CA}{4}\left(F^{\a\b}\dual{F}_{\a\b}\right)P^\nu.
\ee
{Here, $T_{\fl}^{\mu\nu}$ is the fluid energy-momentum tensor (see \cite{Sadooghi2018} for some details on the derivation of \eqref{eq:csmhd}). The inhomogeneous Maxwell equation leads to $\partial_{\mu}\JT^{\mu}=0$. Using the homogeneous Maxwell equation, the $\Theta$ dependent part in $\JT^{\mu}$ trivially vanishes, and we are left with $\partial_{\mu}J^{\mu}=0$. }\par
{As it is argued in \cite{Kharzeev:2009fn}}, the comoving temporal derivative of the axion-like field gives rise to the chiral chemical potential,
\begin{equation}\label{eq:axion-cme}
	\MUF \equiv u^\mu P_\mu,
\end{equation}
while, {according to \cite{Sadooghi2018}}, its comoving spatial gradient produces the \AH\,current\footnote{{See Appendix \ref{app:notation} for the definition of the comoving temporal and spatial derivatives.}}
 {
\be
J_{\text{AH}}^{\mu}=C_{A}\epsilon^{\mu\nu\alpha\beta}E_{\nu}u_{\alpha}P_{\beta}.
\ee
Decomposing $F_{\mu\nu}$ and its dual tensor appearing in \eqref{eq:mcs-current} in terms of the electric and magnetic field according to  \cite{Bekenstein1978}

\bel{eq:f-decomp}
F_{\mu\nu} = E_\mu u_\nu -E_\nu u_\mu - \epsLmnab B^\a u^\b,\qquad \dual{F}_{\mu\nu} = B_\mu u_\nu -B_\nu u_\mu + \epsLmnab E^\a u^\b,
\ee
where $u_{\mu}$ is the fluid velocity, and using  \eqref{eq:axion-cme} the \CM\,current $\CA\MUF B^\mu$, and the AH current
$C_{A}E^{\mu\nu}P_{\nu}$}
emerge as a part of $\JT^\mu$ \cite{Kharzeev:2009fn,Sadooghi2018}. {Here, $E^{\mu\nu}$ is defined as $E^{\mu\nu}\equiv \epsilon^{\mu\nu\alpha\beta}E_{\alpha}u_{\beta}$.}
\subsection{\MCS\,thermodynamics}
Before we deduce the hydrostatic configuration, we need to understand the thermodynamics of the \MCS\,theory. To {this purpose}, we follow the variational method utilized in \cite{Bekenstein2006,Aguiar:2003pp}. {We start with the} effective action of the fluid, the {axion-like field $\Theta$}, and the \EM\,fields\,\cite{Friedman2013}
\begin{equation}
	S = \intx{x}{4}\left[-\EP(s,\ELCH,\axion)+\mcL_{\mcs}\right].
\end{equation}
Here, $\EP$ and $s$ are the energy and entropy densities, and $\ELCH\equiv u_\mu J^\mu$. We assume that the fluid has no other conserved current but $\JT^\mu$. This is a common assumption at high temperatures. In our definition of $\ELCH$, we have replaced $\JT^\mu$ with $J^\mu$ since, {as previously stated,} the divergence of the $\Theta$-dependent part of the former is trivial. The derivatives of $\EP$ {with respect to its variables} are defined as\,\cite{Ozonder:2010zy}
\begin{equation}
	\Big(\pderiv{\EP}{s}\Big)_{\ELCH,\axion} \equiv T,\qquad\Big(\pderiv{\EP}{\ELCH}\Big)_{s,\axion} \equiv \ELMU,\qquad
	\Big(\pderiv{\EP}{\axion}\Big)_{\ELCH,s} \equiv \RAXION.
\end{equation}
{They lead to the first law of thermodynamics,
\bel{eq:firstlaw}
\rd{\EP} = T\rd{s} + \ELMU \rd{\ELCH} + \RAXION \rd{\axion}.
\ee
}
In {a thermal} equilibrium, $T$ and $\ELMU$ are identical to the temperature and chemical potential, respectively.
We apply the variational principle to the action under the following constraints:
\begin{equation}\label{eq:action-constraints}
	u^\mu u_\mu = 1,\qquad \p_\mu J^\mu = 0,\qquad \p_\mu S^\mu =0.
\end{equation}
Here, $S^\mu$ is the entropy current. The first constraint comes from the requirement that $u^\mu$ must be timelike. The second one is a consequence of \eqref{eq:mcs}, and the third one of the conditions of {thermal} equilibrium. For later convenience, we decompose $S^\mu$ and $J^\mu$ {parallel and perpendicular to $u^{\mu}$}
\begin{equation}
	S^\mu = su^\mu + \PROJ^{\mu\nu}S_\nu,\qq{and}J^\mu = \ELCH u^\mu + \PROJ^{\mu\nu}J_\nu,
\end{equation}
{where $\PROJ^{\mu\nu}$ is defined in \rfApp{app:notation}.}
{As in \cite{Bekenstein2006,Aguiar:2003pp}}, we introduce an effective Lagrangian, with Lagrange multipliers $\lambda$, $\xi$, and $w$, that enforces {the} constraints  \eqref{eq:action-constraints}\,
\begin{equation}
	\mcL_{\text{eff.}} = -\EP(s,\ELCH,\axion)+\mcL_{\mcs} + \lambda \p_\mu J^\mu + \xi \p_\mu S^\mu - \half w \left(u^\mu u_\mu - 1\right).
\end{equation}
Integrating by part, the effective Lagrangian is rewritten as
\begin{eqnarray}\label{eq:efflag}
	\mcL_{\text{eff.}} &=& -\EP(s,\ELCH,\axion)-\inv{4}F^{\mu\nu}F_{\mu\nu} - \frac{\CA}{4}\axion F^{\mu\nu}\dual{F}_{\mu\nu} \nn\\ &&-  \left(\ELCH u^\mu + \PROJ^{\mu\nu}J_\nu\right) \left(A_\mu+\p_\mu\lambda\right) - \left(su^\mu + \PROJ^{\mu\nu}S_\nu\right)\p_\mu\xi  - \half w \left(u^\mu u_\mu - 1\right).
\end{eqnarray}
{The} variations of the effective Lagrangian with respect to $u_\mu$, $\axion$, $\ELCH$, and $s$ give rise to
\begin{align}
	&\label{eq:w-by-lm}w u_\mu =-\ELCH (A_\mu + \p_\mu \lambda) - s\p_\mu \xi ,\\
	&\label{eq:raxion}\RAXION = - \frac{\CA}{4}F^{\mu\nu}\dual{F}_{\mu\nu},\\
	&\label{eq:mu-gauge}\ELMU = -u^\mu (A_\mu + \p_\mu \lambda),\\
	&\label{eq:T-by-lm}T = - u^\mu \p_\mu \xi .
\end{align}
We note that \eqref{eq:mu-gauge} is consistent with the results of \cite{Jensen2013} with an overall change of sign due to different metric signature conventions. Plugging \eqref{eq:T-by-lm} and \eqref{eq:mu-gauge} into \eqref{eq:w-by-lm}, we find
\begin{equation}\label{eq:enthalpy}
	w = Ts+\mu \ELCH.
\end{equation}
{Assuming at this stage that the fields which are present in the \CS\,Lagrangian are background fields, the full partition function of the theory is given by}
	\begin{equation}
		Z = \int \mathcal{D}\Phi\exp(\int_0^{\beta}\rd{\tau}\intx{x}{3}\left(\mcL_{\text{CS}}+\mcL[\Phi]\right))
		= \exp(-\f{V}{T}\mcL_{\text{CS}})Z_{\Phi}.
	\end{equation}	
Here, {$\Phi$ stands generically for all the other fluctuating fields, and $Z_{\Phi}$ is their corresponding partition function}. {Focusing particularly on the $\Theta$ dependent part of the Lagrangian, the corresponding thermodynamic potential $\Omega$ is thus given by}
\begin{equation}\label{eq:omega}
	\Omega(T,\ELMU,\axion) = -\frac{T}{V}\ln Z = \mcL_{\text{CS}} + \cdots,
\end{equation}
 gives rise to
\begin{equation}\label{eq:raxion-p}
	\Big(\pderiv{\Omega}{\axion}\Big)_{\ELMU,T}=\RAXION.
\end{equation}
Using $\PR=-\Omega$\,\cite{Kapusta2011}, {the definition of $\mcL_{\text{CS}}$ from \eqref{eq:mcs-lagrangian}, and  the constraint relation \eqref{eq:raxion}, the Gibbs-Duhem relation is respectively modified as} \cite{Aguiar:2003pp,Ozonder:2010zy}
\begin{eqnarray}\label{eq:mcs-thermo}
\rd{\PR} = s\rd{T} + \ELCH \rd{\ELMU} - \RAXION \rd{\axion}.
\end{eqnarray}
{Combining this expression with \eqref{eq:firstlaw},} we recognize $w$ in \eqref{eq:enthalpy} as fluid's specific enthalpy {density}, i.e. $w=\EP+\PR$. According to \eqref{eq:raxion-p}, $w$ is independent of $\axion$, and the ordinary relation for entropy density is {thus} not modified,
\begin{equation}\label{eq:thermo}
	s = \inv{T}\left(\EP + \PR - \ELCH\ELMU \right).
\end{equation}
\subsection{Hydrostatic equilibrium}
\par
We are now in a position to fix the hydrostatic equilibrium state\,\cite{Hiscock:1985zz}. We start from the covariant generalization of the thermodynamic identity\,\eqref{eq:thermo}, which reads \cite{Kovtun:2016lfw,Hernandez:2017mch,Hiscock:1985zz,Israel:1976tn,Becattini:2012tc}
\begin{equation}\label{eq:entropyc}
	S^\mu = \PR \beta^\mu + T_{\fl}^{\mu\nu}\beta_\nu - \FUG J^{\mu},
\end{equation}
where $\FUG\equiv\mu/T$, and $\beta^\mu\equiv u^\mu/T$. Taking the divergence of $S^\mu$ and using \eqref{eq:csmhd} {as well as $\partial_{\mu}J^{\mu}=0$ from \eqref{eq:action-constraints}} gives rise to
\begin{eqnarray}\label{eq:s-div}
	\p_\mu S^\mu &=& \p_\mu \PR  \beta^\mu+ \PR \p_\mu\beta^\mu + \left(F^{\nu\lambda}J_\lambda - \frac{\CA}{4}\left(F^{\a\b}\dual{F}_{\a\b}\right)P^\nu\right)\beta_\nu + T_{\fl}^{\mu\nu}\p_\mu\beta_\nu
-J^{\mu}\p_\mu\FUG
	\nn\\&=&
	 \p_\mu \PR  \beta^\mu+ \PR \p_\mu\beta^\mu  + \half T_{\fl}^{\mu\nu}\left(\p_\mu\beta_\nu+\p_\nu\beta_\mu\right)- \frac{\CA}{4}\left(F^{\a\b}\dual{F}_{\a\b}\right)\beta^\lambda P_{\lambda}
	\nn\\&&-
	J^\mu\left[\frac{E_\mu}{T}+\p_\mu\FUG\right].
\end{eqnarray}
In {thermal} equilibrium, the divergence {of $S^{\mu}$} must vanish. This is obtained if $\beta^\mu$ is the symmetry of the hydrostatic equilibrium state, in the sense that the Lie derivative of every physical quantity vanishes\,\cite{Zee2013}. Hence, $\beta$ is a Killing vector\,\cite{Becattini:2012tc}
\begin{equation}\label{eq:killing}
	\p_\mu\beta_\nu+\p_\nu\beta_\mu = 0.
\end{equation}
This immediately eliminates $T_{\fl}^{\mu\nu}\left(\p_\mu\beta_\nu+\p_\nu\beta_\mu\right)$, and $\p_\mu\beta^\mu$ in \eqref{eq:s-div}. Noting that the Lie derivative of a scalar $\phi$ with respect to $\beta^\mu$ is simply $\beta^\mu \p_\mu \phi$, $\beta^{\lambda}P_{\lambda}=\beta^\lambda \p_\lambda \axion$ also vanishes. { This implies, $\MUF=\partial_{0}\Theta=0$ in LRF of the fluid}.  For the last term {of \eqref{eq:s-div}} to vanish it is sufficient that
\begin{equation}\label{eq:elmu-equib}
	E_\mu=-T\p_\mu\FUG.
\end{equation}
{Although, the $\beta$-symmetry leads to the time-independency of hydrodynamic variables in the LRF of the fluid, their spatial gradients are not necessarily vanishing \cite{Jensen2013}. This is why, we can assume that $\tvec{P}_0=\boldsymbol{\nabla}\Theta$ in  $P_\mu = \left(0,\tvec{P}_0\right)$ does not vanish in equilibrium.} {Let us also notice that} if there is no chemical potential other than $\MUF$ present, the electric four-vector {vanishes} at equilibrium. Hereafter, we assume that this is the case. The general solution to the \eqref{eq:killing} reads\,\cite{DeGroot1980,Becattini:2012tc}
 \begin{equation}
 	\beta_\mu(x) = \frac{u^\mu}{T} + \omega_{\mu\nu}x^\nu,
 \end{equation}
where $\omega_{\mu\nu}$ is the thermal vorticity tensor defined as $\omega_{\mu\nu}\equiv-\half\left(\p_\mu\beta_\nu-\p_\nu\beta_\mu\right)$. {Assuming that} in the hydrostatic configuration the thermal vorticity is zero,  the solution reduces to time-independent temperature and four-velocity.
With the above considerations, for the hydrostatic equilibrium in the \LRF, we have
\begin{eqnarray}\label{eq:hydrostatic}
	\EP &=& \EPEQ,\qquad \PR = \PREQ,\qquad s = s_{0},\qquad T = T_{0},\qquad u^\mu = (1,\tvec{0}),
	\nn\\
	B^\mu &=& \left(0,\tvec{B}_0\right),\qquad E^\mu = 0,\qquad \MUF = 0,\qquad\grad\axion=\tvec{P}_0,\qquad\axion=\axion_0.
\end{eqnarray}
The subscript $0$ is used to denote that the quantities are constants.\footnote{{We note that it is legitimate, and sometimes fruitful\,\cite{Pu:2009fj,Kovtun:2019hdm}, to consider the fluctuations from a moving observer's perspective for which the fluid four-velocity reads $u^\mu = \gamma (1,\tvec{v})$. We come back to this issue in \rfsec{sec:modes-lrf}.}} One can check that the configuration of \eqref{eq:hydrostatic} satisfies \eqref{eq:csmhd}. Hereafter, we assume the following form for $J^\mu$
\begin{equation}\label{eq:ohm}
	J^\mu = \ELCH u^\mu + \ELCOND E^\mu,
\end{equation}
in which $\ELCOND$ is the electric conductivity. {Contracting} the inhomogeneous \MCS\,equation, i.e. the second equation in \eqref{eq:mcs}, {with}  $\beta_\nu$, and using  \eqref{eq:hydrostatic}, we find $u_\mu \JT^{\mu} = 0$. {Using the definition of $\JT^{\mu}$ from \eqref{eq:mcs-current}, with $J^{\mu}$ from \eqref{eq:ohm}, we arrive at}
\begin{equation}
	\ELCHEQ = \CA \tvec{P}_0\cdot\tvec{B}_0.
\end{equation}
The above equation suggests that in \CSMHD, the local charge density can be nonzero in equilibrium. However, a nonvanishing $\ELCHEQ$ requires a corresponding nonvanishing chemical potential {$\mu_e$}. Since we have already assumed that such a chemical potential does not exist, we also need to assume that the local charge density is zero.  By this virtue
\begin{equation}
	\ELCHEQ = 0,\qq{and}\tvec{P}_0\cdot \tvec{B}_0=0.
\end{equation}
We should emphasize that in contrast to \cite{Aguiar:2003pp,Ozonder:2010zy}, our setup is dissipative. Out of equilibrium, the entropy production is governed by the following relation
\begin{equation}
	T\p_\mu S^\mu = \ELCOND E^2.
\end{equation}

\section{Collective modes of the \CSMHD}\label{sec:modes-lrf}
\setcounter{equation}{0}
In this section, we find the collective mode of the \MCS\,theory in the \LRF\,{of the fluid}. To find the collective modes, we introduce perturbations to the hydrostatic configuration of \eqref{eq:hydrostatic}\,\cite{Hiscock:1983zz,Hiscock:1985zz}. We then solve the \EOM, up to first order in perturbations
\begin{eqnarray}\label{eq:peom}
	&&\p_\mu \dual{\delta F}^{\mu\nu} = \order*{\delta^2},\qquad \p_\mu \delta F^{\mu\nu} - \delta \JT^\nu=\order*{\delta^2},\nonumber\\
	&&\p_\mu \delta T_{\fl}^{\mu\nu} - \delta \left(F^{\nu\lambda}J_\lambda\right) + \frac{\CA}{4}\delta\left[\left(F^{\a\b}\dual{F}_{\a\b}\right)P^\nu\right]=\order*{\delta^2}.
\end{eqnarray}
We assume perturbations around the hydrostatic configuration of {the} form
\[\delta \tilde{X}\sim\delta X\exp(-i\omega t+i\tvec{k}\cdot\tvec{x}),\]
for each hydrodyanmic variable $X$. Plugging the perturbed variables in \eqref{eq:peom}, and keeping terms up to the first order in perturbations, gives rise to a system of linear equations as\,\cite{Pu:2009fj}
\begin{equation}\label{eq:secular}
	M \delta X = 0,
\end{equation}
where $M$ is the matrix of coefficients and $\delta X$ the unknown perturbative variables. Equation \eqref{eq:secular} is a polynomial equation whose solutions of form $\omega=\omega(\tvec{k})$ are the so-called modes of theory. The modes are called (non)hydro modes\footnote{(Non)hydro modes are also called (gapped)gapless modes.}, if $\omega\left(\tvec{k}\right)$ is (not) zero for $\tvec{k}=\tvec{0}$. Modes are stable if $\Im(\omega(\tvec{k}))<0$, for all values of $\tvec{k}$, and they are causal if\,\cite{Pu:2009fj,Kovtun:2019hdm}
\begin{equation}\label{eq:asym-caus-condi}
	\lim_{k\to\infty}\abs{\f{\Re(\omega)}{k}}\leq 1\,,
\end{equation}
wherein $k \equiv \sqrt{\tvec{k}.\tvec{k}}$.
\rfl{eq:secular} is not analytically solvable for all modes. To investigate their stability, we thus use the Routh-Hurwitz stability criterion \cite{Routh,Hurwitz1895,Krotscheck1978}.
\par
{Employing the \MCS\,equations, it is possible to reduce the number of $\delta X$ to}
\begin{equation}\label{eq:perturbation-variables}
\delta X = \left(\delta T,\delta u_x,\delta u_y,\delta u_z\right),
\end{equation}
{and,} in particular,
\begin{equation}\label{eq:pert}
	\DP = \frac{w_0}{T_0}\DT,\qquad \DEP = \frac{w_0}{\vs^2T_0}\DT,\qquad \delta\ELCH = 0.
\end{equation}
Here, $\vs$ is the speed of sound, and $w_0=\epsilon_0+p_0$. In \eqref{eq:pert},
the first and second equations arise from \eqref{eq:firstlaw} and \eqref{eq:mcs-thermo}. The last relation corresponds to the fact that a nonvanishing $\ELMUEQ$ is required for the charge density fluctuation to be physically possible. 
 Before we proceed, {it is necessary to understand how the gradient of $\axion$ appears in equilibrium.} {To do this, we remind that} the scale at which the axion changes is much larger than the scale of the fluctuations. {We thus consider  $\tvec{P}_0$ as a constant in the linear analysis. } The same {is also true for} the gauge potential $A_{0,\mu}$ in equilibrium; if one assumes that $A_{0,\mu}$ has nonvanishing second-order derivatives, and hence $\tvec{B}_0$ has nonvanishing gradients, the \Alfven\,and magnetosonic excitations disappear. By this virtue, we let $\axion_0 = \dotpr{\tvec{P_0}}{\xv}$. {Further, assuming that}  $\tvec{B_0}=B_0\uvy$ in the \LRF\!\!, we have
	\begin{eqnarray}\label{eq:fluctuations}
		\axion &=& \dotpr{\tvec{P_0}}{\xv} +  \DAX\,,\qq{with}\tvec{P_0}\equiv\left(P_x,0,P_z\right)\,,\nn\\ u_\mu &=& \left(1,-\tvec{\delta v}\right)\,,\qq{with}\tvec{\delta v}=\left(\delta u_x,\delta u_y,\delta u_z\right)\,,
		\nn\\
		B_\mu &=& B_0\left(\delta u_y, 0,-1,0\right)+\left(0,-\tvec{\delta B}\right)\,,\qq{with}\tvec{\delta B}\equiv\left(\delta B_x,\delta B_y,\delta B_z\right)\,,\nn\\ E&=&\left(0,-\tvec{\delta E}\right)\,,\qq{with}\tvec{\delta E}\equiv\left(\delta E_x,\delta E_y,\delta E_z\right)\,,
		\nn\\
		\tvec{k}&=&k\left(\sin\polar\cos\azim,\cos\polar,\sin\polar\sin\azim\right)\,,\qq{with}k =\sqrt{\tvec{k}.\tvec{k}}\,.
	\end{eqnarray}
	Here, $\polar$ is the polar angle with the zenith direction taken parallel to $\tvec{B_0}$, {and} $\azim$ is the azimuthal angle defined in the $xz$ plane. Plugging \eqref{eq:fluctuations} into the homogeneous \MCS\,equation [first equation in\,\eqref{eq:mcs}], and keeping terms up to the first-order, gives rise to
	\begin{eqnarray}\label{eq:hom-sol}
		\delta B_x &=& -\frac{k}{\omega}\Big(\left(B_0\delta u_x-\delta E_z\right)\cos{\polar}+\delta E_y \sin\azim\sin\polar\Big)\,,\nn\\
		\delta B_y &=& \frac{k\sin\polar}{\omega}\Big(\left(B_0\delta u_x-\delta E_z \right)\cos{\azim}+\left(\delta E_x+B_0\delta u_z\right)\sin\azim\Big)\,,
		\nn\\
		\delta B_z &=& -\frac{k}{\omega}\Big(\left(B_0\delta u_z+\delta E_x\right)\cos{\polar}-\delta E_y \cos\azim\sin\polar\Big)\,.
	\end{eqnarray}
	Plugging the above results into \eqref{eq:mcs-current}, the total current $\JT^\mu$ is found to be
	\begin{eqnarray}\label{eq:j-hom-sol}
		\mcJ_0 &=& \frac{\CA k}{\omega} \Bigg(\Big(\tvec{P_0}\cross\tvec{\delta E}+\left(\tvec{P_0}\cdot\tvec{\delta v}-i\omega\DAX\right)\tvec{B_0}\Big)\cdot\uvkpar+\left(\uvkbot\cross\tvec{P_0}\right)\cdot\tvec{\delta E}\Bigg),
		\nn\\
		\mcJ_x &=& -\ELCOND \delta E_x - \CA \left(\tvec{P_0}\cross\tvec{\delta E}\right)\cdot\uvx\,,
		\nn\\
		\mcJ_y &=& -\ELCOND \delta E_y - \CA \Big(\left(\tvec{P_0}\cross\tvec{\delta E}\right)\cdot\uvy
		+\left(\tvec{P_0}\cdot\tvec{\delta v}-i\omega\DAX\right)
		B_0\Big)\,,
		\nn\\
		\mcJ_z &=& -\ELCOND \delta E_z - \CA \left(\tvec{P_0}\cross\tvec{\delta E}\right)\cdot\uvz\,.
	\end{eqnarray}
	Here, $\uvkbot$ and $\uvkpar$ are the perpendicular and parallel parts of the wavenumber unit vector $\tvec{\hat{k}}\equiv \tvec{k}/k$ with respect to $\tvec{B}_0$,
	\begin{equation*}
		\uvkbot\equiv \left(\cos\azim,0,\sin\azim\right)\sin\polar\,,\qq{and}\uvkpar\equiv \left(0,1,0\right)\cos\polar.
	\end{equation*}
	{As a next step,} we solve the inhomogeneous \MCS\,equation [second equation in\,\eqref{eq:mcs}] to reduce the unknown perturbations to {the four given in} \eqref{eq:perturbation-variables}. The results are too cumbersome to be reproduced here. Plugging {then the resulting expressions into}  the energy-momentum conservation relation \eqref{eq:csmhd}, leads to an equation of form \eqref{eq:secular}. For this equation to be solvable, the determinant of $M$ must vanish. We decompose the determinant into two channels:
	\begin{equation}\label{eq:detm}
		\det(M) = s_0 w_0 \mathcal{D}_{\text{NC}} \mathcal{D}_{\text{C}}\,,
	\end{equation}
	where the channels read
	\begin{eqnarray}
		\label{eq:nc-channel}\ncch &=& B_0^2\ELCOND\left(2\omega^2-\vs^2k^2\right)\left(\omega^2-k^2\right)-2w_0\omega\left(i \omega^2-\ELCOND\omega-i k^2\right)\left(\omega^2-\vs^2k^2\right)\nn\\&&-B_0^2\vs^2k^2\ELCOND\left(\omega^2-k^2\right)\cos(2\polar)\,,\\
		\label{eq:c-channel}\cch&=&8\cos\polar\Bigg(B_0^2\ELCOND k^2\left(\inv{4}-\sin^{2}\polar\right)+\left[w_0\omega\left(i \omega^2-\ELCOND\omega-i k^2\right)-B_0^2\ELCOND\left(\omega^2-\f{3}{4}k^2\right)\right]\Bigg)
		\nn\\
		&&+8\CA \left(w_0+B_0^2\right)\Big(\omega^2-\frac{B_0^2}{B_0^2+w_0}k^2\cos^{2}\polar\Big)\left(\uvkbot\cross\tvec{P_0}\right)\cdot\uvy\,.
	\end{eqnarray}
	The first channel, i.e. $\ncch$, is independent of the \CS\,term in the Lagrangian \eqref{eq:mcs-lagrangian}, and we call it nonchiral. By the same virtue, we refer to the second one, i.e. $\cch$, as the chiral channel. {In particular, $\tvec{P_0}$ appears only in $\cch$. }
	\subsection{\iMHD\,limit}
	Before we proceed, we may benchmark our method with the known results of \iMHD\, in    \cite{Gedalin1993}. To do so, we expand $\det(M)$ from \eqref{eq:detm} in powers of $\ELCOND$, and keep the highest order term. This gives rise to
	\begin{eqnarray}\label{eq:imhd-detm}
		 \Big(\frac{B_0^2}{B_0^2+w_0}&&\left(\omega^2-k^2\right)\left(\omega^2-\vs^2k^2\cos^{2}\polar\right)+\frac{w_0}{B_0^2+w_0}\omega^2\left(\omega^2-\vs^2k^2\right)\Big)\nn\\
		&&\times\Big(\frac{B_0^2}{B_0^2+w_0}\left(\omega^2-k^2\cos^{2}\polar\right)+\frac{w_0}{B_0^2+w_0}\omega^2\Big)=0\,.
	\end{eqnarray}
	The first two eigenfrequencies are the relativistic \Alfven\,modes
	\begin{equation}\label{eq:imhd-Alfven}
		\omega_{\rm A,\pm} = \pm \val k \cos\polar\,,
	\end{equation}
	with  the \Alfven\,speed $\val$, {defined by}
	\begin{equation}\label{eq:Alfven-speed}
		\val^2 \equiv \frac{B_0^2}{B_0^2+w_0}\,.
	\end{equation}
	We identify four other eigenfrequencies as the frequencies of slow ($\omega_{\rm sms,\pm}$) and fast ($\omega_{\rm fms,\pm}$) magnetosonic modes
	\begin{equation}\label{eq:imd-magnetosonic}
		\omega_{\rm sms,\pm} = \pm\vs k\sqrt{\mcA-\mcB}\,,\qquad\omega_{\rm fms,\pm} = \pm\vs k\sqrt{\mcA+\mcB}\,,
	\end{equation}
{with}
	\begin{equation}\label{eq:aux-quantities}
		\mcA \equiv \half\Big(1-\val^2\sin^{2}\polar + \frac{\val^2}{\vs^2}\Big)\,,\qquad \mcB\equiv\sqrt{\mcA^2-\left(\frac{\val\cos\polar}{\vs}\right)^2}.
	\end{equation}
	In the absence of the magnetic field, all eigenfrequencies vanish, {except} those of the fast magnetosonic modes. The latter reduces to the sound mode in the {perfect fluid},
	\begin{equation*}
		\omega_{\rm A,\pm} = 0,\qquad \omega_{\rm sms,\pm} = 0\,\qquad \omega_{\rm fms,\pm} = \pm \vs k\,.	
	\end{equation*}
	Even in the presence of the magnetic field, the \Alfven\,and slow magnetosonic modes do not propagate in the plane perpendicular to the magnetic field, i.e. $\polar=\pi/2$:
	\begin{equation*}
		\omega_{\rm A,\pm} = 0,\qquad \omega_{\rm sms,\pm} = 0,\qquad \omega_{\rm fms,\pm} = \pm \vs k\sqrt{1-\val^2+\left(\frac{\val}{\vs}\right)^2}\,.	
	\end{equation*}
	\subsection{Nonchiral channel}
	{We now analyze} the nonchiral channel. This channel is a polynomial of order five, and according to Abel's impossibility theorem\,\cite{Pesic2004}, an exact solution for the {corresponding} eigenfrequencies cannot be obtained. The nature of these modes can, however, be revealed using a long-wavelength expansion
	\begin{eqnarray}\label{eq:nc-modes}
		\omega_{\text{sms},\pm} &=& \pm\vs k\sqrt{\mcA-\mcB}-\frac{i (1-\val^2)\left(1-\vs^2(\mcA-\mcB)\right)\left(1-\mcA+\mcB\right)}{4\ELCOND\mcB}k^2+\order{k^3},\nn\\
		\omega_{\text{fms},\pm} &=& \pm\vs k\sqrt{\mcA+\mcB}-\frac{i (1-\val^2)\left(1-\vs^2(\mcA+\mcB)\right)\left(\mcA-1+\mcB\right)}{4\ELCOND\mcB}k^2+\order{k^3},\nn\\
		\omega_{\text{NC},\text{nh}} &=& - \frac{i\ELCOND}{1-\val^2}  + \frac{i (1-\val^2)\left(1+\vs^2\left(1-2\mcA\right)\right)}{\ELCOND}k^2+\order{k^3}.
	\end{eqnarray}
	The first two eigenfrequencies belong to the slow and fast magnetosonic {nonchiral (NC)} eigenfrequencies modified by the presence of finite electrical conductivity. As \eqref{eq:nc-modes} suggests, the resistivity $1/\ELCOND$ damps the magnetosonic modes. The third  {NC}  eigenfrequency is genuine and contains a nonhydro part. Due to the positivity of the electrical conductivity, {which is} required by the second law of thermodynamics\,\cite{Kovtun:2012rj}, the nonhydro mode is stable. A thorough examination of {its} linear stability using the Routh-Hurwitz criteria is presented in \rfApp{app:rh-anal}. The analysis proves that the nonchiral channel is linearly stable.
	\par
	{To check whether this channel is causal, we use} the asymptotic causality condition \eqref{eq:asym-caus-condi}. Expanding $\ncch$ in terms of $k$, and keeping the highest order term gives rise to
	\begin{equation}\label{eq:nc-caus}
		\ncch\sim 2i k^5
		\vg\left(\vg^2-\vs^2\right)\left(\vg^2-1\right)w_0\,,
	\end{equation}
{where $v_g$ is the group velocity, which in $k\to \infty$ is given by $v_g\sim \omega/k$}.
	{Setting \eqref{eq:nc-caus} equal to zero, it turns out that}  the {asymptotic} group velocity does not exceed the speed of light. {This indicates that} the chiral channel is causal.

	\subsection{Chiral channel}
	{In what follows, we investigate} the chiral channel.
	{To do this, let us first emphasize} that the eigenfrequencies of this channel vanish in the direction of the magnetic field, i.e. $\polar=\pi/2$. For simplicity, we define
	\begin{equation}\label{eq:p-redef}
		\tvec{P}_0 = P \left(\cos(\phi-\Delta),0,\sin(\phi-\Delta)\right)\quad\implies\quad \left(\uvkbot\cross\tvec{P_0}\right)\cdot\boldsymbol{\hat{y}} = P\sin\Delta\sin\polar\,.
	\end{equation}
	Here, $\Delta$ is the angle between the vector $\tvec{P}_0$ and the transverse wavenumber vector $\uvkbot$. Assuming $0\leq \Delta < 2\pi$, we can constraint $P$ to be positive. Although this channel is polynomial of order three, its exact solution is too complicated to be useful. Therefore, similarly to the nonchiral case, we perform a long-wavelength expansion to obtain
	\begin{eqnarray}\label{eq:c-modes}
		\omega_{\rm A,\pm} &=& \pm \val k \cos\polar - \frac{i (1-\val^2)(1-\val^2\cos^{2}\polar)}{2\left(\ELCOND-\CA P\sin\Delta\tan\polar\right)}k^2\nn\\
		&&
		\pm \frac{\left(1-\val^2\right)^2\left(1-5\val^2\cos^{2}\polar\right)\left(1-\val^2\cos^{2}\polar\right)}{8\val\cos\polar\left(\ELCOND-\CA P\sin\Delta\tan\polar\right)^2}k^3+\order{k^4}
		\,,\nn\\
		\omega_{\rm C,\rm nh} &=& - \frac{i\left(\ELCOND-\CA P\sin\Delta\tan\polar\right)}{1-\val^2}- \frac{i (1-\val^2)(1-\val^2\cos^{2}\polar)}{\left(\ELCOND-\CA P\sin\Delta\tan\polar\right)}k^2+\order{k^3}\,.
	\end{eqnarray}
	{A comparison with \eqref{eq:nc-modes},} suggests the definition of an effective conductivity which mixes the Ohmic and the \AH\,conductivities,
	\begin{equation}\label{eq:sigameff}
		\ELCONDEFF \equiv \ELCOND - \CA P\sin\Delta\tan\polar\,.
	\end{equation}
Let us assume $0\leq \polar < \pi/2$. For $0\leq\Delta<\pi$, the \AH\,current is in the opposite direction of the Ohm one. Since $\tan\polar$ is unbounded, for any value of $P$, there exists a critical value for $\polar$ such that for polar angles larger than that $\ELCONDEFF$ becomes negative. Consequently, the nonhydro mode $\omega_{\rm C,\rm nh}$ becomes unstable, and the \Alfven\,modes $\omega_{\rm A,\pm}$ are amplified. On the other hand, for $\pi\leq\Delta<2\pi$, the \AH\,current enhances the Ohm current. Therefore, the nonhydro mode is stable, and the \Alfven\,modes are damped.
	The same remarks hold for $\pi/2< \polar \leq \pi$, with modes in $0\leq\Delta<\pi$ interval being stable and the ones in $\pi\leq\Delta<2\pi$ being stable. In summary, a mode propagating in some angle $\polar$ might be stable while its mirrored one in the opposite angle is unstable. This is a manifestation of parity symmetry breaking {caused by the CP violating Chern-Simons term}. The Routh-Hurwitz analysis in \rfApp{app:rh-anal} confirms these remarks.
	\par
	\begin{figure*}[!hbt]\centering
		\includegraphics[width=.6\linewidth]{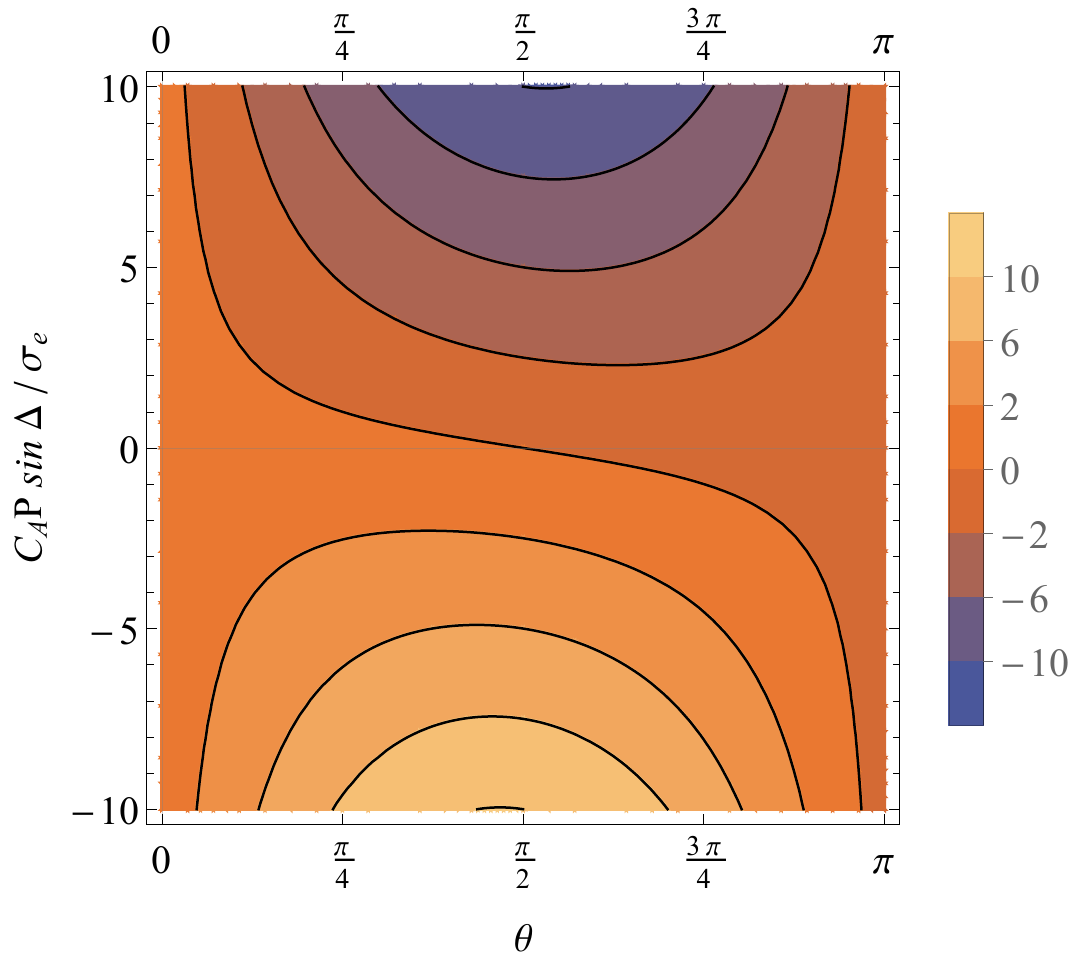}
		\caption{Depiction of $\ELCONDEFF\cos\polar/\ELCOND$. The horizontal axes is the polar angle $\polar$, and the vertical one is $\CA P \sin\Delta / \ELCOND$. As the figure suggests, for any value of $P$ there exists a range of $\polar$ and $\Delta$ for which $\ELCONDEFF\cos\polar<0$ and the chiral channel {becomes} unstable.}\label{fig:effsigma}
	\end{figure*}
	As in the nonchiral case, we check the causality of the chiral channel using \eqref{eq:asym-caus-condi}. The leading term for the short wavelength expansion reads
	\begin{equation}\label{eq:c-caus}
		\cch\sim 8i k^3 \vg\left(\vg^2-1\right)w_0\,,
	\end{equation}
which implies that the chiral channel is causal. The causality of the chiral channel has crucial consequences. First, it implies that the linear stability of the system in a moving frame is similar to the \LRF.\footnote{{This is confirmed by explicit computations, which are not reproduced in the present work.}} Second, \AH\,instability is not fictitious. One may be tempted to write a relaxation equation for the current to remove the instability. However, such an approach does not work. {Let us remind that the relaxation time approach is essentially employed to avoid instantaneous propagation of signals.  The \CSMHD\,modes are, however, causal. Hence a relaxation time approach seems to be useless. We nevertheless use the following ansatz to check whether it can cure the instability problem of this mode
	\begin{equation}\label{eq:relaxN}
		\tau_J\Delta^{\mu\nu}\tder{J}_\nu + J^\mu = \ELCOND E^{\mu}\,.
	\end{equation}
Let us note that since the axionic part of the current is dissipationless, a relaxation equation can only be written for the Ohmic part of the current. An explicit computation of the modes using \eqref{eq:relaxN} confirms that the AH instability cannot be removed by this relation.}
	\section{Numerical results}\label{sec:numerics}
	\setcounter{equation}{0}
	\par		
	\begin{figure}
		\centering
		\includegraphics[width=.55\linewidth, height=0.45\linewidth]{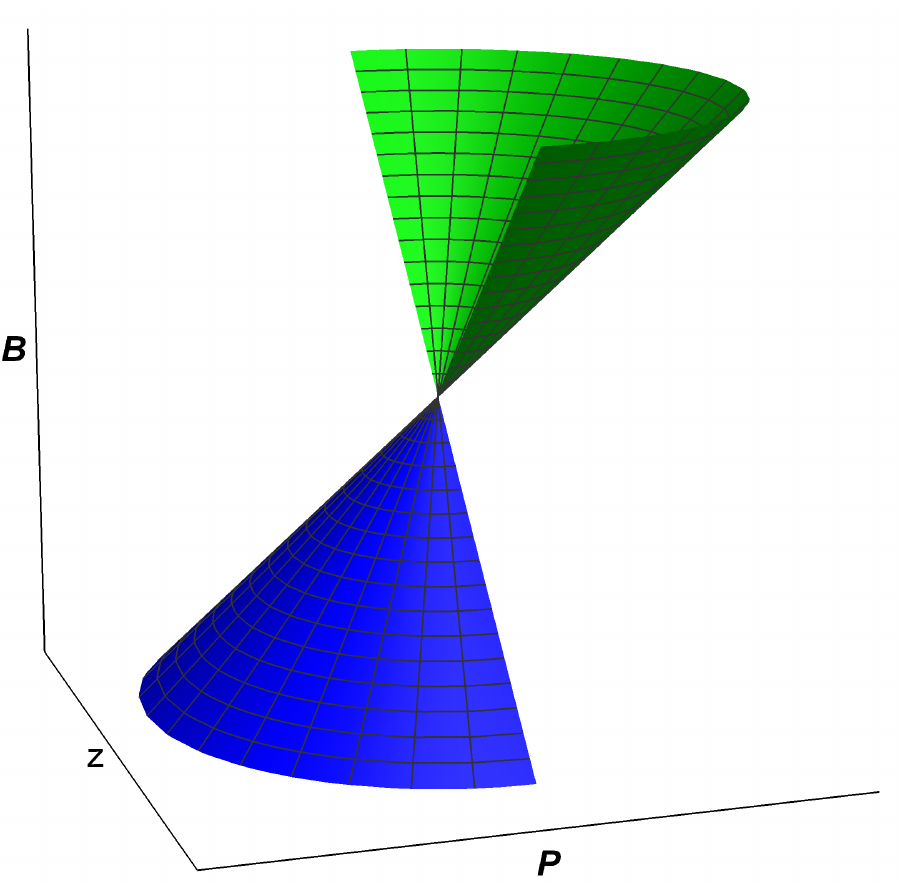}
		\caption{The schematic view of the unstable regions in space. The cone is defined as $\polar=\polar_c$, wherein $\polar_c$ is the angle at which the effective conductivity \eqref{eq:sigameff} vanishes. The upper quarter is  the interval $\polar_c\leq \polar < \pi/2$ (green), while the lower one is $3\pi/2< \polar \leq \pi-\polar_c$ (blue). The chiral channel is unstable within these two quarters and stable outside them.}
		\label{fig:cones}
	\end{figure}
	\begin{figure}
		\centering
		\includegraphics[width=1\linewidth]{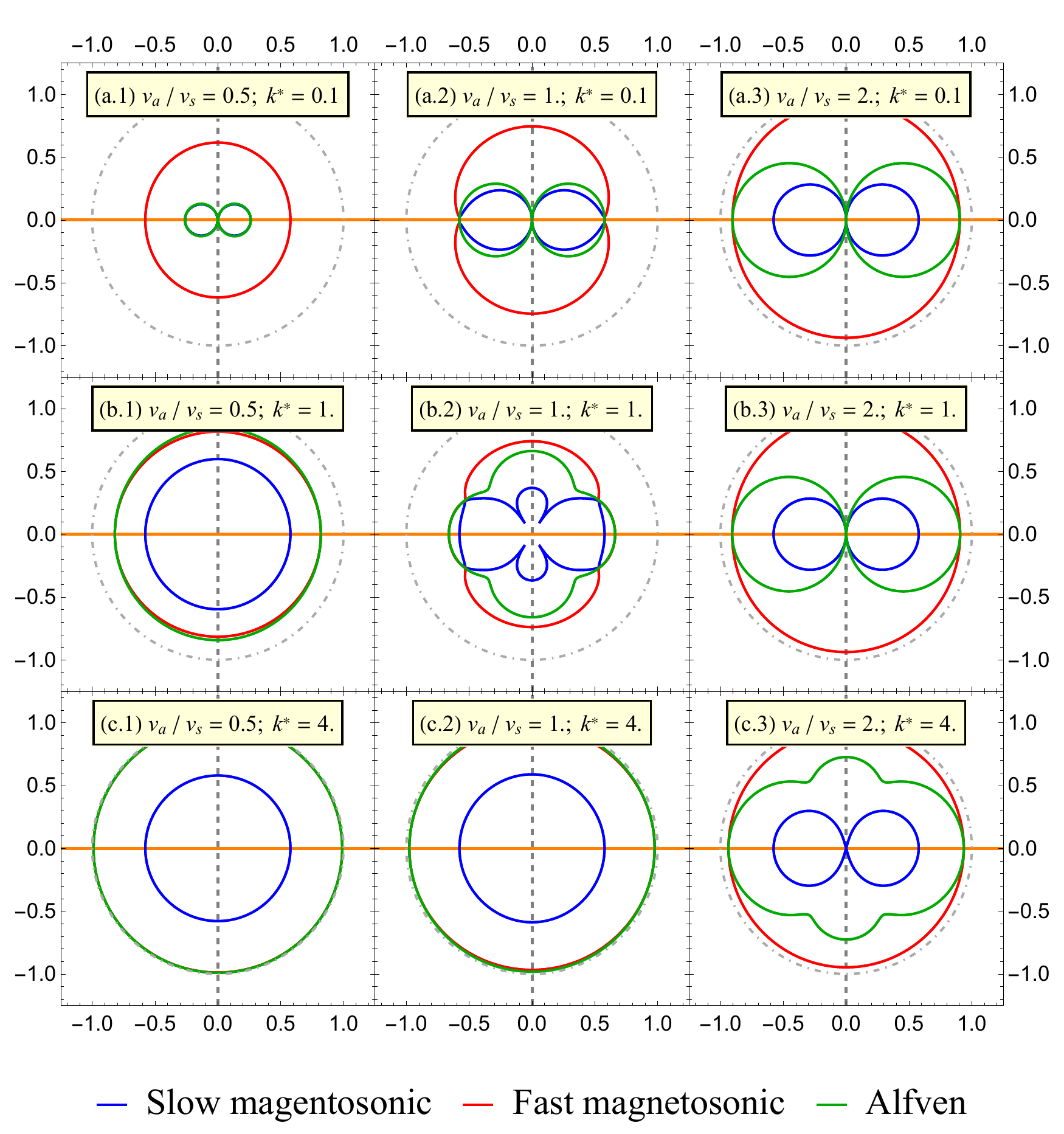}
		\caption{The figures represent the phase velocity $\Re(\omega)/k$ in the plane of $\tvec{B}_0$-$\tvec{P}_0$. The figures are for different values of $k/\ELCOND$ (columns 1, 2, and 3) and $\val/\vs$ (rows a, b, and c). For all figures, $P=\ELCOND/\left(\sqrt{3}\CA\right)$. The horizontal orange gridline in each figure corresponds to the direction of $\tvec{B}_0$, while the vertical dashed gray gridline is in the direction of $\tvec{P}_0$. The dash-dotted circle demonstrates the speed of light. The blue and red curves correspond to slow and fast magnetosonic modes in \eqref{eq:nc-modes}, respectively. The green curves are the phase velocity of the \Alfven\,modes in \eqref{eq:c-modes}. The \Alfven\,modes are symmetric. }
		\label{fig:vak-phase-bp}
	\end{figure} 
	\begin{figure}
		\centering
		\includegraphics[width=1\linewidth]{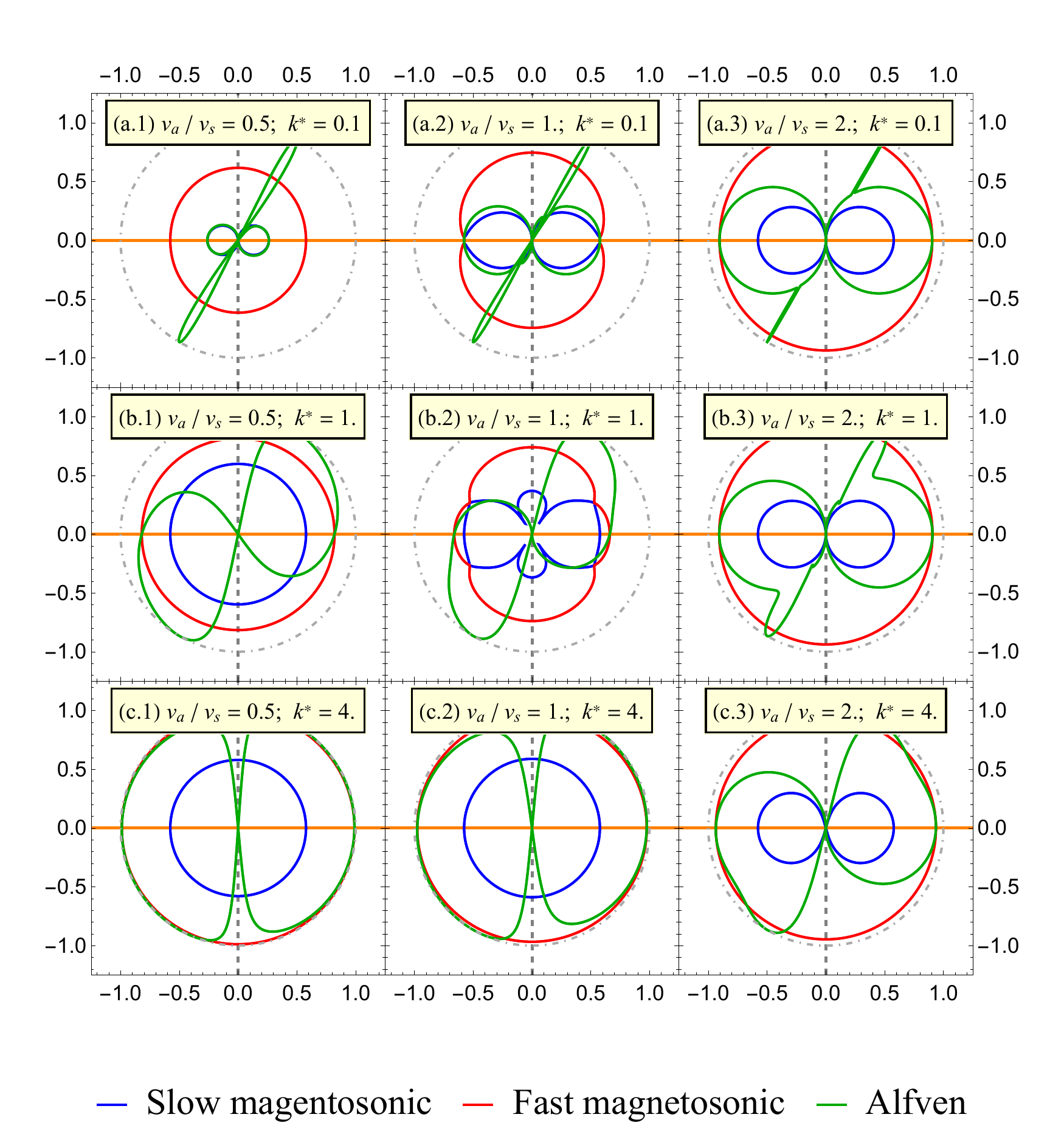}
		\caption{The figures represent the phase velocity $\Re(\omega)/k$ in the plane perpendicular to $\tvec{P}_0$. For the upper half-plane, $\Delta=\pi/2$, and in the lower one $\Delta=3\pi/2$. The figures are for different values of $k/\ELCOND$ (columns 1, 2, and 3) and $\val/\vs$ (rows a, b, and c). For all figures $P=\ELCOND/\left(\sqrt{3}\CA\right)$. The horizontal orange gridline in each figure corresponds to the direction of $\tvec{B}_0$, while the vertical dashed gray gridline is in the direction perpendicular to both $\tvec{B}_0$ and $\tvec{P}_0$. The dash-dotted circle demonstrates the speed of light. The blue and red curves correspond to slow and fast magnetosonic modes in \eqref{eq:nc-modes}, respectively. The green curves are the phase velocity of the \Alfven\,modes in \eqref{eq:c-modes}. Even for small values of $\val/\vs$, the \Alfven\,modes reach the speed of light in a certain direction.}
		\label{fig:vak-phase-bz}
	\end{figure}

	\begin{figure}
		\centering
		\includegraphics[width=.85\linewidth]{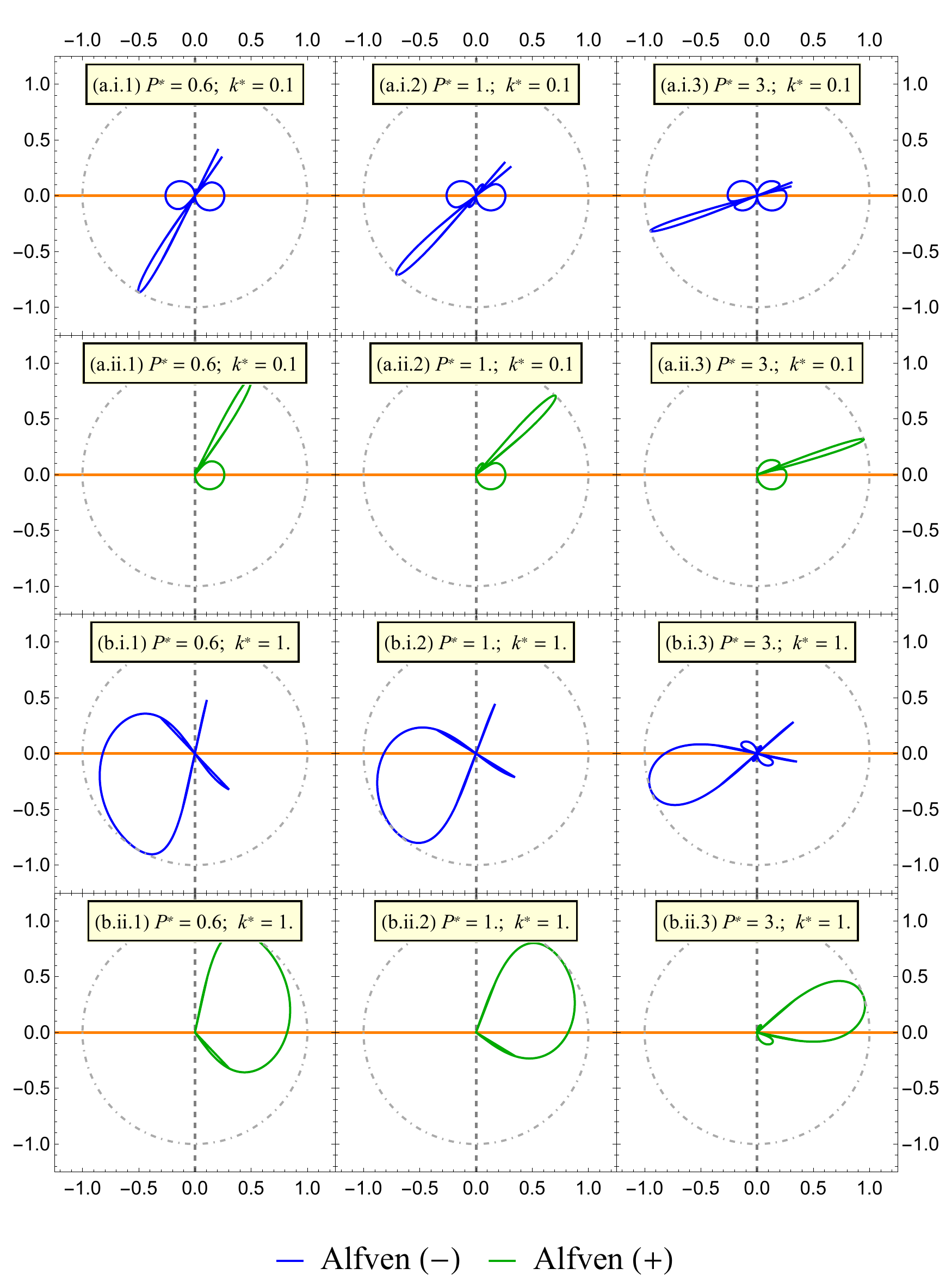}
		\caption{The figures represent the phase velocity $\Re(\omega)/k$ of the \Alfven\,modes in the plane perpendicular to $\tvec{P}_0$. For the upper half-plane, $\Delta=\pi/2$, and in the lower one $\Delta=3\pi/2$. The figures are for different values of $k/\ELCOND$ (columns 1, 2, and 3) and $P^\star=\CA P/\ELCOND$ (rows a and b). Each row is divided into I and ii parts, with i (ii) demonstrating the negative (positive) \Alfven\,modes of \eqref{eq:c-modes}. $\val/\vs=0.5$ for all figures. The horizontal orange gridline in each figure corresponds to the direction of $\tvec{B}_0$, while the vertical dashed gray gridline is in the direction perpendicular to both $\tvec{B}_0$ and $\tvec{P}_0$. The dash-dotted circle demonstrates the speed of light. The positive and negative \Alfven\,modes propagate with the speed of light in opposite polar directions. This direction gets closer to the direction of $\tvec{B}_0$ for larger values of $P^\star$. The asymmetry decreases for larger values of $k/\ELCOND$.}
		\label{fig:kp-phase-bz}
	\end{figure}
\begin{figure}
	\centering
	\includegraphics[width=.85\linewidth]{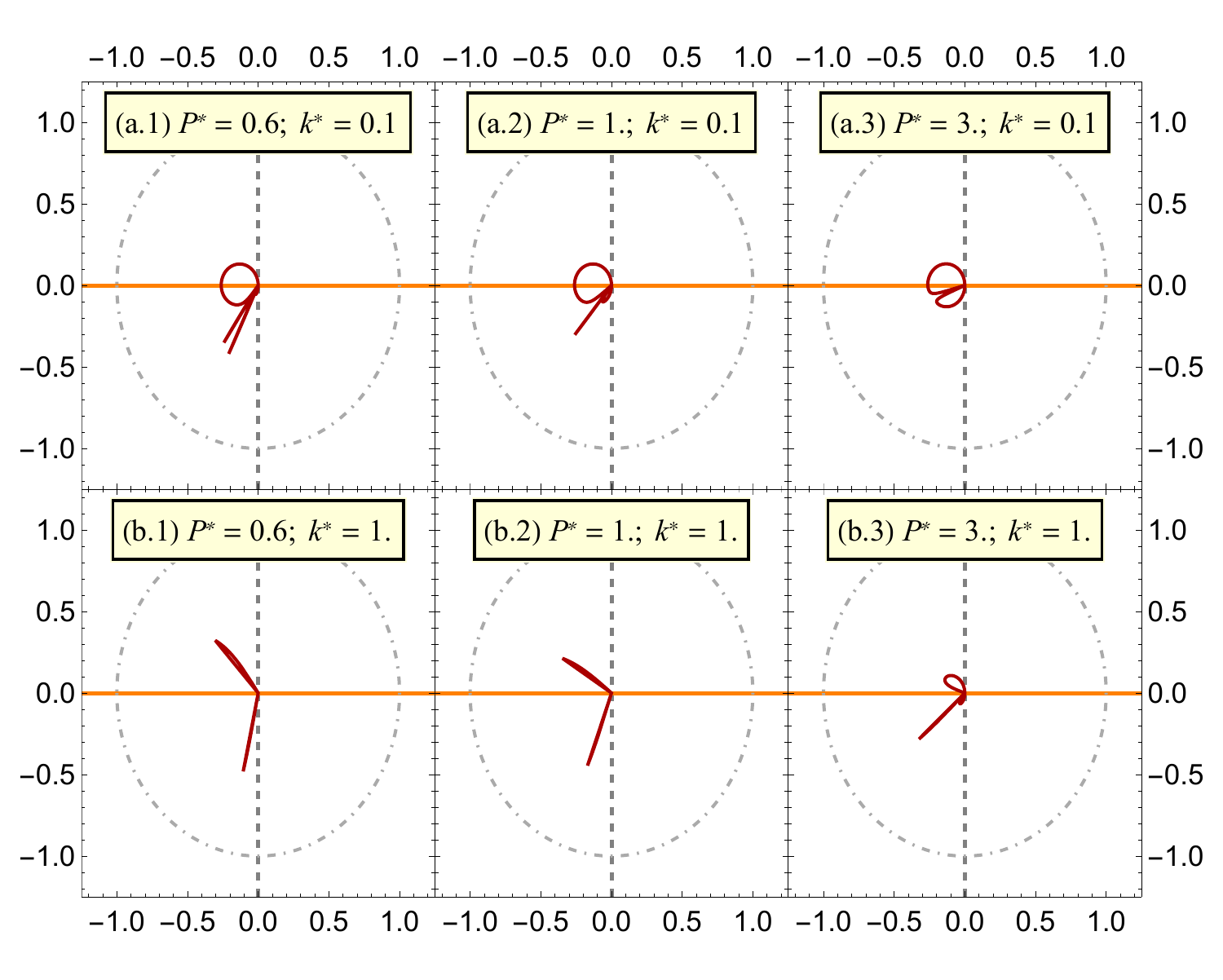}
	\caption{The figures represent the phase velocity $\Re(\omega)/k$ of the chiral nonhydro mode in the plane perpendicular to $\tvec{P}_0$. For the upper half-plane, $\Delta=\pi/2$, and in the lower one $\Delta=3\pi/2$. The figures are for different values of $k/\ELCOND$ (columns 1, 2, and 3) and $P^\star=\CA P/\ELCOND$ (rows a and b).  For all figures, $\val/\vs=0.5$. The horizontal orange gridline in each figure corresponds to the direction of $\tvec{B}_0$, while the vertical dashed gray gridline is in the direction perpendicular to both $\tvec{B}_0$ and $\tvec{P}_0$. The dash-dotted circle demonstrates the speed of light. }
	\label{fig:kp-phase-nh-bz}
\end{figure}
	\begin{figure}
		\centering
		\includegraphics[width=.85\linewidth]{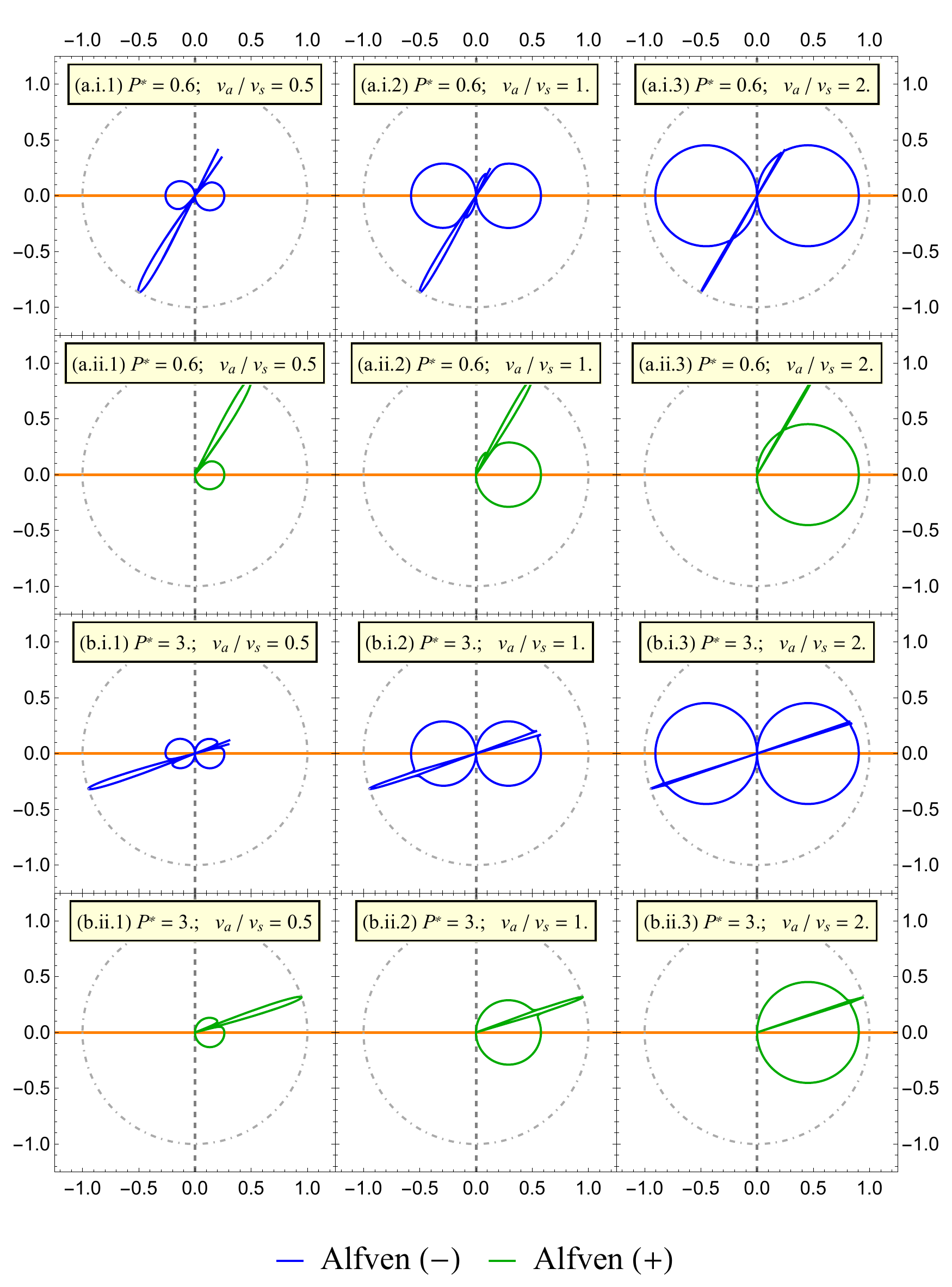}
		\caption{The figures represent the phase velocity $\Re(\omega)/k$ of the \Alfven\,modes in the plane perpendicular to $\tvec{P}_0$. For the upper half-plane, $\Delta=\pi/2$, and in the lower one $\Delta=3\pi/2$. The figures are for different values of $\val/\vs$ (columns 1, 2, and 3) and $P^\star=\CA P/\ELCOND$ (rows a and b). Each row is divided into I and ii parts, with i (ii) demonstrating the negative (positive) \Alfven\,modes of \eqref{eq:c-modes}. $k/\ELCOND=0.1$ for all figures. The horizontal orange gridline in each figure corresponds to the direction of $\tvec{B}_0$, while the vertical dashed gray gridline is in the direction perpendicular to both $\tvec{B}_0$ and $\tvec{P}_0$. The dash-dotted circle demonstrates the speed of light. The positive and negative \Alfven\,modes propagate with the speed of light in opposite polar directions. This direction gets closer to the direction of $\tvec{B}_0$ for larger values of $P^\star$.}
		\label{fig:vap-phase-bz}
	\end{figure}
\begin{figure}
	\centering
	\includegraphics[width=1\linewidth]{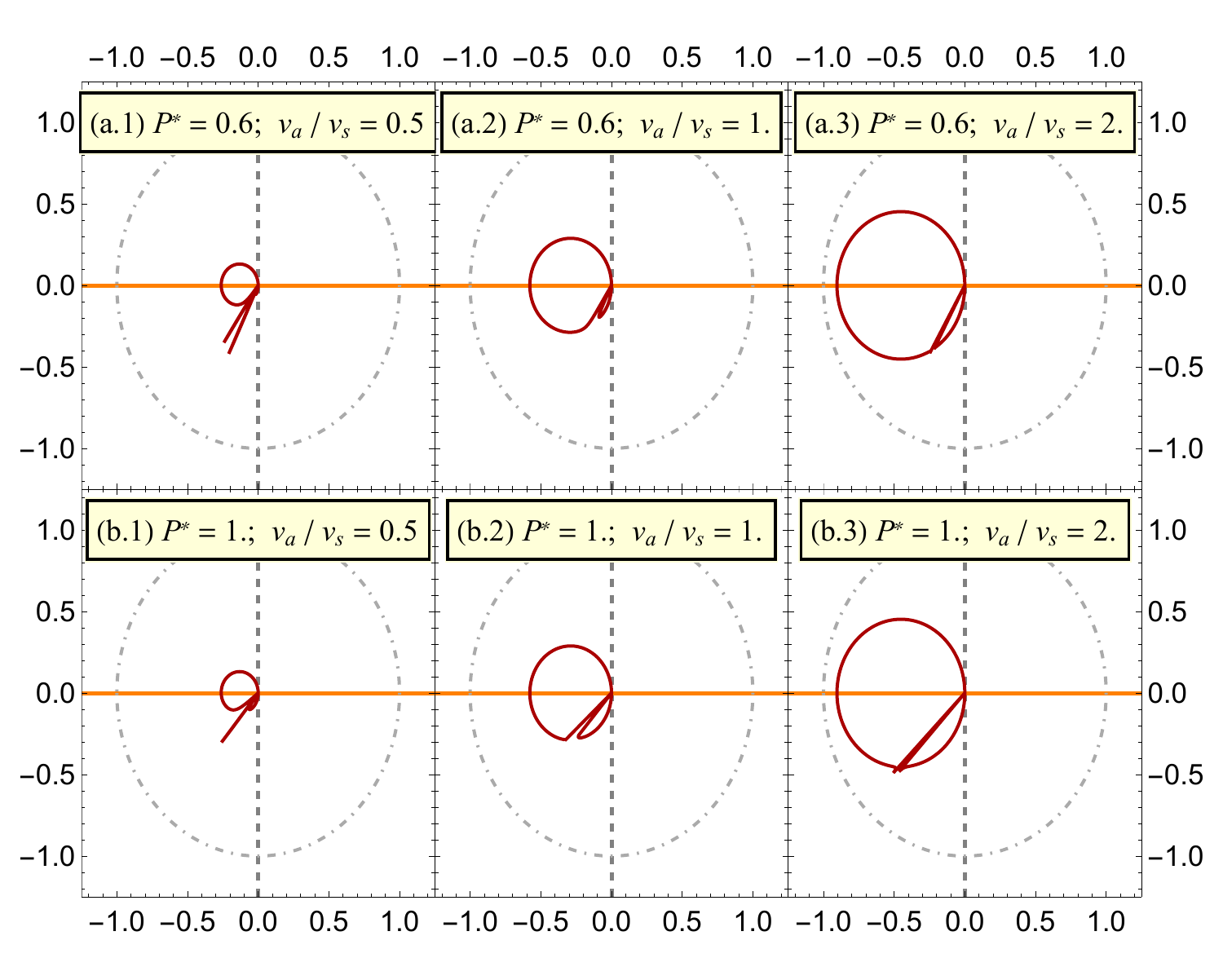}
	\caption{The figures represent the phase velocity $\Re(\omega)/k$ of the chiral nonhydro mode in the plane perpendicular to $\tvec{P}_0$. For the upper half-plane, $\Delta=\pi/2$, and in the lower one $\Delta=3\pi/2$. The figures are for different values of $\val/\vs$ (columns 1, 2, and 3) and $P^\star=\CA P/\ELCOND$ (rows a, and b).  $k/\ELCOND=0.1$ for all figures. The horizontal orange gridline in each figure corresponds to the direction of $\tvec{B}_0$, while the vertical dashed gray gridline is in the direction perpendicular to both $\tvec{B}_0$ and $\tvec{P}_0$. The dash-dotted circle demonstrates the speed of light.}
	\label{fig:vap-phase-nh-bz}
\end{figure}
	\begin{figure}
		\centering
		\includegraphics[width=1\linewidth]{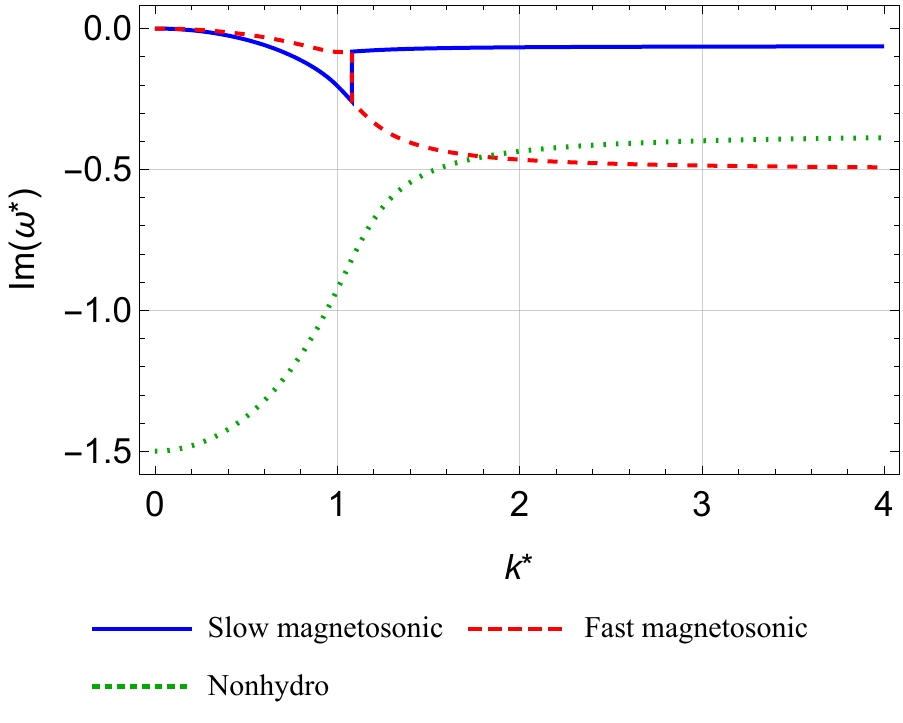}
		\caption{The figure represents $\Im(\omega)/\ELCOND$ vs. $k/\ELCOND$ for the nonchiral channel \eqref{eq:nc-channel}. As shown in the text, $\Im(\omega)$ is always negative and the nonchiral channel is stable. The fast magnetosonic modes are suppressed more strongly than the slow ones. $\Im(\omega)$  becomes almost constant after a sufficiently large value of $k$.  }
		\label{fig:ncim}
	\end{figure}
	\begin{figure}
		\centering
		\includegraphics[width=1\linewidth]{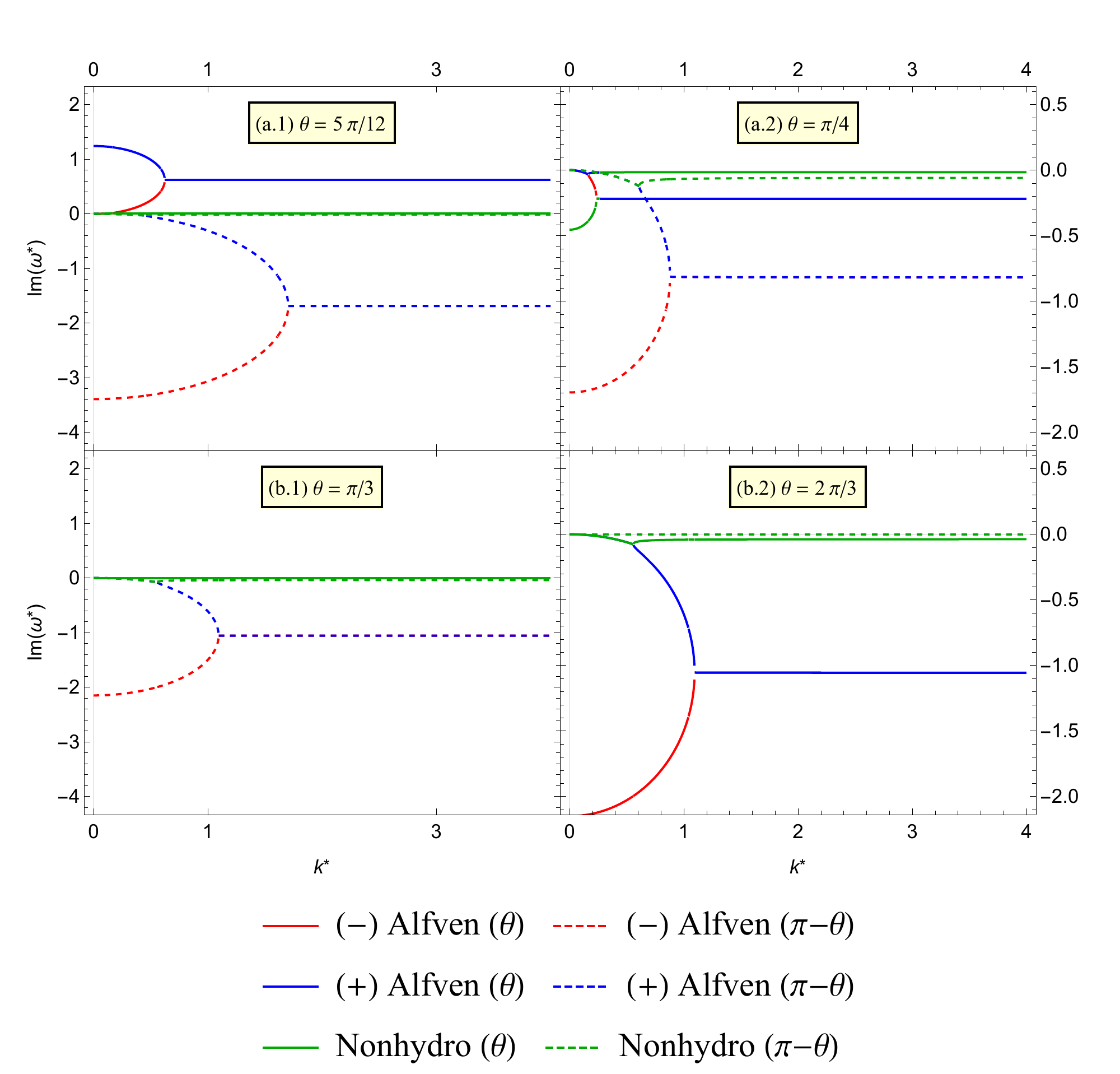}
		\caption{The figure represents $\Im(\omega)/\ELCOND$ vs. $k/\ELCOND$ for the chiral channel \eqref{eq:c-channel}: (a.1) represents a value of $\polar$ within the upper unstable  {region} of \rff{fig:cones}; (a.2) represents a value of $\polar$ within the upper stable {region} of \rff{fig:cones}; (b.1) represents  the angle $\polar_c$ at which $\ELCONDEFF=0$ within the upper region of \rff{fig:cones}; (b.2) represents  the angle $\polar_c$ at which $\ELCONDEFF=0$ within the lower region of \rff{fig:cones}. The mode propagating in exactly opposite direction are drawn with the same color but dashed lines. The channel is still stable at this critical angle. The $\Im(\omega)$  becomes almost constant after a sufficiently large value of $k^{\star}$.}
		\label{fig:impart-chiral}
	\end{figure}
	In this section, we present numerical results for the collective modes. First, we depict the phase velocities $v_{\rm ph} \equiv \Re(\omega/k)$ for different modes. We then plot the imaginary part of eigenfrequencies for the nonchiral and chiral channels. Since our results are independent of the electrical conductivity, we make certain quantities dimensionless by dividing them by $\ELCOND$
	\bel{eq:dimless}
	\omega^\star \equiv \omega/\ELCOND\,,\qquad
	k^\star \equiv k/\ELCOND\,,\qquad P^\star\equiv\CA P/\ELCOND.
	\ee
	{As it turns out,} for any particular choice of parameters, there exists two critical polar angles $\polar_c$ and $\pi-\polar_c$ for which the effective conductivity vanishes
	\bel{eq:polarc}
	\polar_c \equiv \arctan(\frac{1}{P^\star\sin\Delta})\,.
	\ee
These angles divide the space into stable and unstable regions. A schematic picture of this division is presented in \rff{fig:cones}. The chiral channel is unstable inside the green upper ($\polar_c< \polar < \pi/2$) and blue lower ($3\pi/2< \polar < \pi-\polar_c$) quarters. The remained symmetries of the space allow us to choose a particular vertical slice, which is the $\tvec{B}$-$z$ plane, and a particular horizontal one, which is the $\tvec{B}$-$\tvec{P}$ plane. In the $\tvec{B}$-$z$ plane, $\Delta=\pi/2~(3\pi/2)$ for the upper (lower) half. In the $\tvec{B}$-$\tvec{P}$ plane $\Delta=0$ for both halves. We note that the absolute value of $\Delta$ is not significant in our analysis, because it can be absorbed into $P^\star$, but the sign of $\sin\Delta$ matters. For simplicity, we call the different modes of \eqref{eq:c-modes} negative Alfv\'en ($\omega_{\rm A,-}$), positive Alfv\'en ($\omega_{\rm A,+}$), and chiral nonhydro ($\omega_{\rm C,\rm nh}$) {modes}.
	\subsubsection*{\textit{i) Phase velocities}}
We use polar plots to depict the phase velocities. To do so, we need to transform from spherical coordinates to polar coordinates in $\tvec{B}$-$z$ and $\tvec{B}$-$\tvec{P}$ planes. In the $\tvec{B}$-$z$, we define the polar angle as
	\[
	\varphi \equiv {\rm sgn}\left(\sin\Delta\right) \polar\,.
	\]
	 Positive (negative) $\varphi$ corresponds to the upper (lower) half of the $\tvec{B}$-$z$ plane. Since $\Delta=0$ for the $\tvec{B}$-$\tvec{P}$ plane, the upper and lower half-planes are similar. In each figure of Figs.~\ref{fig:vak-phase-bp}-\ref{fig:vap-phase-nh-bz},
the absolute value of the phase velocity at any particular {value of $\varphi$} is equal to the radius of the corresponding curve. The sign of the velocity is not shown, and the sign of plot ticks are just indicators of the corresponding quarter. The phase velocities in the $\tvec{B}$-$\tvec{P}$ plane are depicted in \rff{fig:vak-phase-bp}. The modes behave similarly to those of the resistive \MHD~\cite{Kawazura:2017lpc}. For small values of $k^\star$, the slow magnetosonic and
	 \Alfven\,modes have similar phase velocities, and the fast magnetosonic modes are the fastest ones. This behavior is not surprising because the limit $k\ll\ELCOND$ is the \iMHD\,limit. As $k^\star$ increases, the \Alfven\,modes obtain phase velocities closer to the fast magnetosonic ones. In the nonchiral channel, the phase velocity vanishes for $\cos\polar=0$, as it can be analytically found from \eqref{eq:nc-channel}. On the other hand, a similar general statement cannot be expressed for the phase velocity at $\sin\polar=0$. The special case of $k^\star=1$, for which the phase velocity of slow magnetosonic modes becomes equal to the speed of sound,  is interesting. The nonhydro mode of the nonchiral channel does not propagate, i.e. its phase velocity is always zero.
The modes in the $\tvec{B}$-$\tvec{P}$ plane are symmetric, and we do not represent any other figure in this plane.
	 \par
We represent the phase velocities in the $\tvec{B}$-$z$ plane with some details. In \rff{fig:vak-phase-bz}, the phase velocities of all modes are drawn. In the stable polar region, the phase velocities are similar to those of the $\tvec{B}$-$\tvec{P}$ plane. However, in the unstable region, the \Alfven\,modes behave drastically different. Even for small values of $k^\star$, there exists a region for which the \Alfven\,modes propagate with the speed of light. We can understand this behavior by inspecting the first $k$-dependent term in \Alfven\,phase velocity which is found from \eqref{eq:c-modes},
	\bel{eq:alfven-vp}
	v_{\rm ph,\Alfven} = \pm \val\cos\polar
	\pm \frac{\left(1-\val^2\right)^2\left(1-5\val^2\cos^{2}\polar\right)\left(1-\val^2\cos^{2}\polar\right)}{8\val\cos\polar\ELCONDEFF^2}k^2 + \order{k^3}.
	\ee
	When $\ELCONDEFF$ becomes very small, the phase velocity increases. But one should keep in mind that the higher order terms are absent in \eqref{eq:alfven-vp}, and the phase velocity does not actually tend to infinity as this relation suggests. Also, as \eqref{eq:alfven-vp} suggests, this region widens as $k^\star$ increases. For sufficiently large $k^\star$, the \Alfven\,modes obtain the speed of light. In contrast to the magnetosonic modes, the \Alfven\,ones are asymmetric under the mirror symmetry with respect {to the direction of} $\tvec{B}$: The chiral channel is not symmetric under transformation of $\polar\to \pi-\polar$, while the nonchiral one is. Increasing $\val/\vs$, which for a fixed temperature corresponds to stronger magnetic fields, has the same effect as in $\tvec{B}$-$\tvec{P}$ plane.

	In the nonchiral channel, the nonhydro mode has nonzero phase velocity. The negative and positive \Alfven\,modes overlap with each other and the nonhydro one. Therefore, to better understand the behavior of the chiral channel, we draw the phase velocities separately. The phase velocity of the nonhydro mode cannot be understood using the long-wave expansion \eqref{eq:alfven-vp}. However, we can rely on numerical inspection to understand the peculiar behavior of the chiral channel. We start with the chiral channel's modes in the upper half-plane. In the first stable region, namely $0<\polar<\polar_c$, only the \Alfven\,modes propagate. The phase velocities have opposite signs but equal values. The velocities of these hydro modes are enhanced by increasing $k^\star$ (\rff{fig:kp-phase-bz}) and $\val/\vs$ (\rff{fig:vap-phase-bz}). The velocities tend to the speed of light as we get closer to the critical angle. At the critical angle, the negative \Alfven\,mode is replaced by the nonhydro one.  Both modes propagate with the speed of light. For the nonhydro mode, this only happens exactly at the critical angle and is not captured in Fig.  \rff{fig:kp-phase-nh-bz}. Then we enter the upper unstable region, i.e. $\polar_c<\polar<\pi/2$. In this region, there is a subregion for which the modes propagate. The subregion widens in larger values of $k^\star$. For small values of $k^\star$, the negative \Alfven\,mode obtain positive velocities, while the positive \Alfven\,mode is replaced by the nonhydro mode, which has a negative velocity. As $k^\star$ is increased, the nonhydro mode is suppressed, and the positive and negative \Alfven\,modes obtain velocities with the \textit{right} signs. The same remarks hold for the second quarter of the upper half-plane. The modes behave similarly in the lower half-planes, with negative and positive \Alfven\,modes swapped.
	\subsubsection*{\textit{ii) Imaginary parts}}
	{In Figs. \ref{fig:ncim} and \ref{fig:impart-chiral} the $k^{\star}$ dependence of the imaginary parts of all modes in both channels are plotted. As it turns out, they} become almost constant after a particular value of $k^\star$. {In particular,} the imaginary parts of the nonchiral channel modes, presented in  \rff{fig:ncim}, are always negative. {This confirms our proof presented in \rfApp{app:rh-anal}.} For the chiral channel, the imaginary parts become positive in the unstable regions. Let us also notice that in \rff{fig:impart-chiral} the modes in the exact opposite direction have negative imaginary parts. The imaginary parts of the modes at the two critical angles are also depicted in \rff{fig:impart-chiral}. All imaginary parts vanish in the direction of the critical angle, while in the exact opposite direction they have nonvanishing negative values. Therefore at the critical angle, the chiral channel is still stable.
	\section{Concluding remarks}\label{sec:conc}
	\setcounter{equation}{0}
	\par								
	{In the present work,} we performed an analysis of the linear stability of a resistive \CSMHD.  We started with the \MCS\,Lagrangian that produces the {CM current}  through the comoving temporal derivative of an axion-like field. After reproducing the results of \cite{Ozonder:2010zy} for the \MCS\,thermodynamics, we identified the global equilibrium state in \CSMHD\,by applying {a standard} entropy current analysis. We showed that the axial chemical potential $\MUF$ vanishes in global equilibrium, but the spatial gradient of the axion $\tvec{P}=\grad{\axion}$ can give rise to a nonzero electrical charge density. To proceed, we chose the conjugate chemical potential of the electrical charge density $\ELMU$ to be zero in the equilibrium. This choice is equivalent to the power counting scheme in which the magnetic field is of order $\order{1}$, while the electric field is of order $\order{\partial}$. Hence, {in this weak electric field regime \cite{Hernandez:2017mch}, the electric field vanishes in the thermodynamical equilibrium.}  As a consequence, the electric charge density vanishes, and the spatial gradient of the axion-like field is constrained to be perpendicular to the magnetic field. With the hydrostatic configuration fixed, we introduced linear perturbation to find the collective modes. We found that there exist three extra modes in \CSMHD, in addition to the six ones of \iMHD. These nine modes are divided into two channels: Five in a nonchiral or non-axionic channel and four in a chiral or axionic one. The nonchiral channel consists of slow and fast magnetosonic modes, which are damped by the nonzero electrical resistivity. This channel also possesses a nonhydro (gapped) mode. Using the Routh-Hurwitz criteria and asymptotic causality condition, we showed that the nonchiral channel is linearly stable and causal. The chiral channel includes the modified \Alfven\,modes and a nonhydro (gapless) mode. The stability of these modes is controlled by a combination of the Ohm and \AH\,conductivities, which can be considered as a {novel} effective conductivity. In contrast to the Ohm conductivity, effective conductivity becomes negative for the modes propagating sufficiently close to the direction of the magnetic field. Consequently, the chiral channel is unstable in {this} region. However, this channel is causal, and therefore the instability is physical. We {also} performed a numerical inspection of phase velocities and imaginary parts of different modes. As our results show,  there is a critical angle that separates stable and unstable regions of the space for the chiral channel. In the direction of this critical angle, the \Alfven\,waves travel with the speed of light without becoming unstable.
	\par
	The current work has {a} theoretical nature, {in which we explored the stability and causality of the modes  propagating in a chiral medium}.  Although the \CM\,current that arises from the \MCS\,theory has a physical explanation, this theory is not the only approach to the \CME. To the best of our knowledge, other consequences of the \MCS\,theory are not well understood in the context of the \QGP\,physics. In particular, in contrast to the condensed matter physics\,\cite{Liu2016}, we are unaware of a physical explanation for the occurrence of the \AH\,effect in the \QGP.\footnote{{In \cite{Sadooghi2018}, we have presented another application of the presence of the \AH\,current within \CSMHD}.}  It might be interesting to investigate the possible mechanisms that give rise to the \AH\,effect in different states of strongly interacting {quark} matter.
	\par
	We close this paper by suggesting two possible directions that extend this work. In the present work, we have assumed that the electric chemical potential is zero, which is equivalent to the  assumption of electric field being of order $\order{\partial}$. A possible extension would be to consider the strong electric field regime, in which the electric field is of order $\order{1}$ and the electric chemical potential is nonzero in the equilibrium. Another interesting extension is to assume the equilibrium state to be in a rigid rotation. The work in both directions is in progress.
	\appendix
\section{Notations, conventions and useful formulae}\label{app:notation}
\setcounter{equation}{0}
The energy-momentum tensor of the perfect fluid is {given by}  \cite{Landau1987Fluid,rezzolla,Kovtun:2012rj}
\bel{eq:ideal-fluid}
T^{\mu\nu}_{\fl(0)}  = \ed u^\mu u^\nu - \PR \PROJ^{\mu\nu}.
\ee
Here, $\ed$ is the energy density, $\PR$ the pressure, and $u^\mu$ the fluid four-velocity normalized as $u^\mu u_\mu = 1$. These so-called hydrodynamic variables have unique definitions for the perfect fluid\,\cite{Kovtun:2012rj}. Consequently, the \LRF\,is unambiguously defined by $u^\mu = (1,\tvec{0})$. In \eqref{eq:ideal-fluid}, $\PROJ_{\mu\nu}\equiv g_{\mu\nu}-u_\mu u_\nu$ projects vectors and tensors in the direction orthogonal to $u_\mu$. The comoving temporal $\tder{}$ and spatial derivatives $\cdbot{\mu}{}$ read
\bel{eq:cm-der}
\tder{} \equiv u^\mu \p_\mu,\qquad \cdbot{\mu}{} \equiv \PROJ^{\nu}_\mu \p_\nu.
\ee
As any antisymmetric tensor of rank two, $F_{\mu\nu}$ and $\dual{F}_{\mu\nu}$ can be decomposed with respect to the timelike vector $u_\mu$\,\cite{Bekenstein1978}
{\be
F_{\mu\nu} = E_\mu u_\nu -E_\nu u_\mu - \epsLmnab B^\a u^\b,\qquad \dual{F}_{\mu\nu} = B_\mu u_\nu -B_\nu u_\mu + \epsLmnab E^\a u^\b,
\ee}
where the \EM\,four-vectors are defined as
\bel{eq:em-fv}
E^\mu \equiv F^{\mu\nu}u_\nu,\qquad B^\mu \equiv \half \epsUmnab F_{\nu\a} u_\b.
\ee
One should bear in mind that only for the comoving observer, say in the \LRF, the above four-vectors coincide with the physical electric and magnetic fields, i.e. $E^\mu = (0,\tvec{E})$ and $B^\mu = (0,\tvec{B})$. By this virtue
\bel{eq:em-mags}
E \equiv \sqrt{-E^\mu E_\mu} = \abs*{\tvec{E}}_\lrf,\qq{and}B \equiv \sqrt{-B^\mu B_\mu} = \abs*{\tvec{B}}_\lrf.
\ee
We note that while $E$ and $B$ are Lorentz invariants, $\abs*{\tvec{E}}$ and $\abs*{\tvec{B}}$ are not.
	\section{Routh-Hurwitz stability analysis}\label{app:rh-anal}
	\setcounter{equation}{0}
	In this appendix, we apply the Routh-Hurwitz stability criteria {\cite{Krotscheck1978}} to channels found in \rfsec{sec:modes-lrf}.
	\subsection{Nonchiral channel}\label{app:rh-anal-chiral}
	For simplicity, we rewrite the nonchiral channel \eqref{eq:nc-channel} as
	\begin{eqnarray}\label{eq:ncchp}
		\ncch'&\equiv&\frac{1-\val^2}{w_0}\ncch = \ELCOND \val^2 (\omega^2-k^2)\left(2\omega ^2-\vs^2 k^2\right)-2 \omega  (1-\val^2)\left(\omega ^2-\vs^2 k^2\right) \left(i k^2+\omega  (\ELCOND-i \omega )\right)\nn\\
		&&-\vs^2 k^2 \ELCH \val^2 \cos(2\polar)\left(\omega^2-k^2\right).
	\end{eqnarray}
	To apply the Routh-Hurwitz criteria, we perform the substitution $\omega\rightarrow i\zeta $\,\cite{Kovtun:2019hdm}. Consequently, $\ncch'$ is transformed into a $5$th order polynomial in $\zeta$,
	\begin{equation*}
		\ncch' = \sum_{i=0}^{5} a_{i}\zeta^{i}\,.
	\end{equation*}
	{We} employ, at this stage, the Routh-Hurwitz criteria to find {whether}  the real part of $\zeta$ is positive. The Routh table reads
	\begin{equation}\label{eq:nc-routh}
		{\rm R}_{\rm NC}=\left(
		\begin{array}{ccc}
			a_{5}       & a_{3}       & a_{1} \\
			a_{4}       & a_{2}       & a_{0} \\
			b_{1}       & b_{2}       & 0     \\
			c_{1}       & c_{2}=a_{0} & 0     \\
			d_{1}       & 0           & 0     \\
			e_{1}=a_{0} & 0           & 0     \\
		\end{array}
		\right).
	\end{equation}
	The coefficients $a_i$ read
	\begin{eqnarray*}
		a_{0}&=&2\vs^2 k^4 \ELCOND \val^2\cos^{2}\polar\,,\qquad a_{1}=2 \vs^2(1-\val^2) k^4\,,\qquad a_{2}=2k^2\ELCOND\left[\vs^2+\val^2\left(1-\vs^2\sin^{2}\polar\right)\right]\,,\nn\\
		a_{3}&=& 2 \left(1+\vs^2\right)\left(1-\val^2\right) k^2\,, \qquad a_{4} = 2\ELCOND\,,\qquad a_{5} = 2 \left(1-\val^2\right).
	\end{eqnarray*}
	All of the above coefficients are positive. Therefore, according to the criteria, all other elements in the first column of the Routh table \eqref{eq:nc-routh} must also be positive to ensure $\Re(\zeta)>0$. The next two coefficients are
	\begin{eqnarray*}
		b_{1}&=&\frac{a_{3} a_{4}-a_{2} a_{5}}{a_{4}} = 2k^2\left(1-\val^2\right)\left(1-\val^2+\val^2\vs^2\sin^{2}\polar\right)\,,\nn\\
		b_{2}&=&\frac{a_{4} a_{1}-a_{0} a_{5}}{a_{4}} = 2k^4\vs^2\left(1-\val^2\right)\left(1-\val^2\cos^{2}\polar\right).
	\end{eqnarray*}
	The positivity of the above coefficients is obvious. We now turn to $c_{1}$,
	\begin{eqnarray}\label{eq:nc-c}
		c_{1}&=&\frac{a_{2} b_{1}-a_{4} b_{2}}{b_{1}}= \frac{2k^2\ELCOND}{1-\val^2+\vs^2\val^2\sin^{2}\polar}\Bigg[\left(1-\val^2\right)\left(1-2\vs^2\sin^{2}\polar\right)\nn\\
&&\hspace{3cm}+\vs^4\sin^{2}\polar\left(1-\val^2\sin^{2}\polar\right)\Bigg]\,.
	\end{eqnarray}
	The positivity of the terms outside the brackets is apparent. The expression inside the bracket can be assumed as a second-order polynomial in $\vs^2$, whose discriminant is negative
	\begin{equation*}
		\Delta = -(1-\val^2)\sin[2](2\polar) < 0\,.
	\end{equation*}
Since $(1-\val^2)>0$, $c_1$ is also positive. By the same virtue, {$d_1$ is positive}
	\begin{eqnarray}
		d_{1} &=&\frac{b_{2} c_{1}-b_{1} c_{2}}{c_{1}}=\frac{b_{2} c_{1}-b_{1} a_{0}}{c_{1}}=2\vs^2\left(1-\vs^2\right)^2\left(1-\val^2\right)^2k^4\cos[4](\polar)\Bigg[\left(1-\val^2\right)\left(1-2\vs^2\sin^{2}\polar\right)\nn\\&&\hspace{4cm}+\vs^4\sin^{2}\polar\left(1-\val^2\sin^{2}\polar\right)\Bigg]^{-1}>0\,.
	\end{eqnarray}
	We conclude that all elements of the first column of the Routh table \eqref{eq:nc-routh} have the same sign.  The nonchiral channel is {thus} stable.
	\subsection{Chiral channel}
	As for the nonchiral case, we rewrite the chiral channel as
	\begin{eqnarray}\label{eq:cchp}
		\cch'=\frac{1-\val^2}{8w_0} &=&  -\ELCONDEFF\cos\polar\left(\omega^2-k^2\val^2\cos^{2}\polar\right)+i \omega\cos\polar\left(1-\val^2\right)\left(\omega^2-k^2\right).
	\end{eqnarray}
	Performing substitution $\omega\rightarrow i \zeta$ gives rise to a third order polynomial in $\zeta$
	\begin{equation*}
		\cch' = \sum_{i=0}^{3} a_{c,i}\zeta^{i}\,,
	\end{equation*}
	The coefficients read
	\begin{equation}\label{eq:cca}
		a_{c,0} = k^2\val^2\ELCONDEFF\cos^{3}\polar\,,\quad a_{c,1} = k^2\left(1-\val^2\right)\cos\polar\,,\quad a_{c,2} = \ELCONDEFF\cos\polar\,,\quad a_{c,3} = \left(1-\val^2\right)\cos\polar\,.
	\end{equation}
	We do not need to reproduce the whole Routh table to realize that the chiral channel is unstable for regions of $\polar$. For the coefficients to have the same sign, it is required that
	\begin{equation*}
		\ELCONDEFF\cos\polar > 0\,.
	\end{equation*}
	As stated in \rfsec{sec:modes-lrf}, such a condition cannot be satisfied for all values of $\polar$. This is visualized in \rff{fig:effsigma}. We conclude that the modes of the chiral channel are always unstable within a region around the direction transverse to the magnetic field.
	\acknowledgments
This work is supported by Sharif University of Technology's Office of Vice President for Research under
Grant No: G960212/Sadooghi. In particular, M. K. thanks this office for financial support. M. S. thanks D. Rischke for valuable discussions.
	\medskip
	\bibliographystyle{JHEP}
	\bibliography{main}

\providecommand{\href}[2]{#2}\begingroup\raggedright\begin{thebibliography}{10}

\bibitem{Landau1987Fluid}
L.D.~Landau and E.M.~Lifshitz, \emph{Fluid Mechanics}, Butterworth-Heinemann,
  2~ed. (Jan., 1987).

\bibitem{rezzolla}
L.~Rezzolla and O.~Zanotti, \emph{Relativistic Hydrodynamics}, OUP Oxford
  (2013).

\bibitem{Schenke2010}
B.~Schenke, S.~Jeon and C.~Gale, \emph{(3+1)d hydrodynamic simulation of
  relativistic heavy-ion collisions},
  \href{https://doi.org/10.1103/PhysRevC.82.014903}{\emph{Phys. Rev. C}
  {\bfseries 82} (2010) 014903}
  [\href{https://arxiv.org/abs/1004.1408}{{\ttfamily 1004.1408}}].

\bibitem{Schenke2011}
B.~Schenke, S.~Jeon and C.~Gale, \emph{Elliptic and triangular flow in
  event-by-event (3+1)d viscous hydrodynamics},
  \href{https://doi.org/10.1103/PhysRevLett.106.042301}{\emph{Phys. Rev. Lett.}
  {\bfseries 106} (2011) 042301}
  [\href{https://arxiv.org/abs/1009.3244}{{\ttfamily 1009.3244}}].

\bibitem{Paquet:2015lta}
J.-F.~Paquet, C.~Shen, G.S.~Denicol, M.~Luzum, B.~Schenke, S.~Jeon et~al.,
  \emph{{Production of photons in relativistic heavy-ion collisions}},
  \href{https://doi.org/10.1103/PhysRevC.93.044906}{\emph{Phys. Rev. C}
  {\bfseries 93} (2016) 044906}
  [\href{https://arxiv.org/abs/1509.06738}{{\ttfamily 1509.06738}}].

\bibitem{Florkowski:2017olj}
W.~Florkowski, M.P.~Heller and M.~Spalinski, \emph{{New theories of
  relativistic hydrodynamics in the LHC era}},
  \href{https://doi.org/10.1088/1361-6633/aaa091}{\emph{Rept. Prog. Phys.}
  {\bfseries 81} (2018) 046001}
  [\href{https://arxiv.org/abs/1707.02282}{{\ttfamily 1707.02282}}].

\bibitem{Hiscock:1983zz}
W.~Hiscock and L.~Lindblom, \emph{{Stability and causality in dissipative
  relativistic fluids}},
  \href{https://doi.org/10.1016/0003-4916(83)90288-9}{\emph{Annals Phys.}
  {\bfseries 151} (1983) 466}.

\bibitem{Hiscock:1985zz}
W.A.~Hiscock and L.~Lindblom, \emph{{Generic instabilities in first-order
  dissipative relativistic fluid theories}},
  \href{https://doi.org/10.1103/PhysRevD.31.725}{\emph{Phys. Rev. D} {\bfseries
  31} (1985) 725}.

\bibitem{Kovtun:2019hdm}
P.~Kovtun, \emph{{First-order relativistic hydrodynamics is stable}},
  \href{https://doi.org/10.1007/JHEP10(2019)034}{\emph{JHEP} {\bfseries 10}
  (2019) 034} [\href{https://arxiv.org/abs/1907.08191}{{\ttfamily
  1907.08191}}].

\bibitem{Romatschke:2016hle}
P.~Romatschke, \emph{{Do nuclear collisions create a locally equilibrated
  quark\textendash{}gluon plasma?}},
  \href{https://doi.org/10.1140/epjc/s10052-016-4567-x}{\emph{Eur. Phys. J. C}
  {\bfseries 77} (2017) 21} [\href{https://arxiv.org/abs/1609.02820}{{\ttfamily
  1609.02820}}].

\bibitem{Heller:2015dha}
M.P.~Heller and M.~Spalinski, \emph{{Hydrodynamics beyond the gradient
  expansion: Resurgence and resummation}},
  \href{https://doi.org/10.1103/PhysRevLett.115.072501}{\emph{Phys. Rev. Lett.}
  {\bfseries 115} (2015) 072501}
  [\href{https://arxiv.org/abs/1503.07514}{{\ttfamily 1503.07514}}].

\bibitem{Kovtun:2012rj}
P.~Kovtun, \emph{{Lectures on hydrodynamic fluctuations in relativistic
  theories}}, \href{https://doi.org/10.1088/1751-8113/45/47/473001}{\emph{J.
  Phys. A} {\bfseries 45} (2012) 473001}
  [\href{https://arxiv.org/abs/1205.5040}{{\ttfamily 1205.5040}}].

\bibitem{Bekenstein1978}
J.D.~Bekenstein and E.~Oron, \emph{{New conservation laws in
  general-relativistic magnetohydrodynamics}},
  \href{https://doi.org/10.1103/PhysRevD.18.1809}{\emph{Phys. Rev. D}
  {\bfseries 18} (1978) 1809}.

\bibitem{Huang:2015oca}
X.-G.~Huang, \emph{{Electromagnetic fields and anomalous transports in
  heavy-ion collisions --- A pedagogical review}},
  \href{https://doi.org/10.1088/0034-4885/79/7/076302}{\emph{Rept. Prog. Phys.}
  {\bfseries 79} (2016) 076302}
  [\href{https://arxiv.org/abs/1509.04073}{{\ttfamily 1509.04073}}].

\bibitem{Denicol:2019iyh}
G.S.~Denicol, E.~Moln\'ar, H.~Niemi and D.H.~Rischke, \emph{{Resistive
  dissipative magnetohydrodynamics from the Boltzmann-Vlasov equation}},
  \href{https://doi.org/10.1103/PhysRevD.99.056017}{\emph{Phys. Rev. D}
  {\bfseries 99} (2019) 056017}
  [\href{https://arxiv.org/abs/1902.01699}{{\ttfamily 1902.01699}}].

\bibitem{Newman:2005hd}
G.M.~Newman, \emph{{Anomalous hydrodynamics}},
  \href{https://doi.org/10.1088/1126-6708/2006/01/158}{\emph{JHEP} {\bfseries
  01} (2006) 158} [\href{https://arxiv.org/abs/hep-ph/0511236}{{\ttfamily
  hep-ph/0511236}}].

\bibitem{Son2009}
D.T.~Son and P.~Surowka, \emph{Hydrodynamics with triangle anomalies},
  \href{https://doi.org/10.1103/PhysRevLett.103.191601}{\emph{Phys. Rev. Lett.}
  {\bfseries 103} (2009) 191601}
  [\href{https://arxiv.org/abs/0906.5044}{{\ttfamily 0906.5044}}].

\bibitem{Florkowski:2017ruc}
W.~Florkowski, B.~Friman, A.~Jaiswal and E.~Speranza, \emph{{Relativistic fluid
  dynamics with spin}},
  \href{https://doi.org/10.1103/PhysRevC.97.041901}{\emph{Phys. Rev. C}
  {\bfseries 97} (2018) 041901}
  [\href{https://arxiv.org/abs/1705.00587}{{\ttfamily 1705.00587}}].

\bibitem{Sadooghi:2016ljd}
N.~Sadooghi and S.M.A.~Tabatabaee, \emph{{The effect of magnetization and
  electric polarization on the anomalous transport coefficients of a chiral
  fluid}}, \href{https://doi.org/10.1088/1367-2630/aa6729}{\emph{New J. Phys.}
  {\bfseries 19} (2017) 053014}
  [\href{https://arxiv.org/abs/1612.02212}{{\ttfamily 1612.02212}}].

\bibitem{Kharzeev:2007jp}
D.E.~Kharzeev, L.D.~McLerran and H.J.~Warringa, \emph{{The effects of
  topological charge change in heavy ion collisions: 'Event by event P and CP
  violation'}},
  \href{https://doi.org/10.1016/j.nuclphysa.2008.02.298}{\emph{Nucl. Phys. A}
  {\bfseries 803} (2008) 227}
  [\href{https://arxiv.org/abs/0711.0950}{{\ttfamily 0711.0950}}].

\bibitem{Fukushima:2008xe}
K.~Fukushima, D.E.~Kharzeev and H.J.~Warringa, \emph{{The chiral magnetic
  effect}}, \href{https://doi.org/10.1103/PhysRevD.78.074033}{\emph{Phys. Rev.
  D} {\bfseries 78} (2008) 074033}
  [\href{https://arxiv.org/abs/0808.3382}{{\ttfamily 0808.3382}}].

\bibitem{Hattori:2017usa}
K.~Hattori, Y.~Hirono, H.-U.~Yee and Y.~Yin, \emph{{MagnetoHydrodynamics with
  chiral anomaly: Phases of collective excitations and instabilities}},
  \href{https://doi.org/10.1103/PhysRevD.100.065023}{\emph{Phys. Rev. D}
  {\bfseries 100} (2019) 065023}
  [\href{https://arxiv.org/abs/1711.08450}{{\ttfamily 1711.08450}}].

\bibitem{Boyarsky2015}
A.~Boyarsky, J.~Fr\"ohlich and O.~Ruchayskiy, \emph{Magnetohydrodynamics of
  chiral relativistic fluids},
  \href{https://doi.org/10.1103/PhysRevD.92.043004}{\emph{Phys. Rev. D}
  {\bfseries 92} (2015) 043004}.

\bibitem{Rogachevskii:2017uyc}
I.~Rogachevskii, O.~Ruchayskiy, A.~Boyarsky, J.~Fr\"ohlich, N.~Kleeorin,
  A.~Brandenburg et~al., \emph{{Laminar and turbulent dynamos in chiral
  magnetohydrodynamics-I: Theory}},
  \href{https://doi.org/10.3847/1538-4357/aa886b}{\emph{Astrophys. J.}
  {\bfseries 846} (2017) 153}
  [\href{https://arxiv.org/abs/1705.00378}{{\ttfamily 1705.00378}}].

\bibitem{Shovkovy:2018tks}
I.~Shovkovy, D.~Rybalka and E.~Gorbar, \emph{{The overdamped chiral magnetic
  wave}}, \href{https://doi.org/10.22323/1.336.0029}{\emph{PoS} {\bfseries
  Confinement2018} (2018) 029}
  [\href{https://arxiv.org/abs/1811.10635}{{\ttfamily 1811.10635}}].

\bibitem{Sadooghi2018}
N.~Sadooghi and M.~Shokri, \emph{{Rotating solutions of nonideal transverse
  Chern-Simons magnetohydrodynamics}},
  \href{https://doi.org/10.1103/PhysRevD.98.076011}{\emph{Phys. Rev. D}
  {\bfseries 98} (2018) 076011}
  [\href{https://arxiv.org/abs/1806.06652}{{\ttfamily 1806.06652}}].

\bibitem{Kharzeev:2009fn}
D.E.~Kharzeev, \emph{{Topologically induced local P and CP violation in QCD
  $\times$ QED}}, \href{https://doi.org/10.1016/j.aop.2009.11.002}{\emph{Annals
  Phys.} {\bfseries 325} (2010) 205}
  [\href{https://arxiv.org/abs/0911.3715}{{\ttfamily 0911.3715}}].

\bibitem{Witten:1979ey}
E.~Witten, \emph{{Dyons of Charge e theta/2 pi}},
  \href{https://doi.org/10.1016/0370-2693(79)90838-4}{\emph{Phys. Lett. B}
  {\bfseries 86} (1979) 283}.

\bibitem{Wilczek1987}
F.~Wilczek, \emph{{Two applications of axion electrodynamics}},
  \href{https://doi.org/10.1103/PhysRevLett.58.1799}{\emph{Phys. Rev. Lett.}
  {\bfseries 58} (1987) 1799}.

\bibitem{Carroll:1989vb}
S.M.~Carroll, G.B.~Field and R.~Jackiw, \emph{{Limits on a Lorentz and parity
  violating modification of electrodynamics}},
  \href{https://doi.org/10.1103/PhysRevD.41.1231}{\emph{Phys. Rev. D}
  {\bfseries 41} (1990) 1231}.

\bibitem{Landsteiner:2016led}
K.~Landsteiner, \emph{{Notes on anomaly induced transport}},
  \href{https://doi.org/10.5506/APhysPolB.47.2617}{\emph{Acta Phys. Polon. B}
  {\bfseries 47} (2016) 2617}
  [\href{https://arxiv.org/abs/1610.04413}{{\ttfamily 1610.04413}}].

\bibitem{Ozonder:2010zy}
S.~Ozonder, \emph{{Maxwell-Chern-Simons hydrodynamics for the chiral magnetic
  effect}}, \href{https://doi.org/10.1103/PhysRevC.81.062201}{\emph{Phys. Rev.
  C} {\bfseries 81} (2010) 062201}
  [\href{https://arxiv.org/abs/1004.3883}{{\ttfamily 1004.3883}}].

\bibitem{Ozonder-Erratum}
S.~Ozonder, \emph{Erratum: Maxwell-chern-simons hydrodynamics for the chiral
  magnetic effect [phys. rev. c 81, 062201(r) (2010)]},
  \href{https://doi.org/10.1103/PhysRevC.84.019903}{\emph{Phys. Rev. C}
  {\bfseries 84} (2011) 019903}.

\bibitem{Kovtun:2016lfw}
P.~Kovtun, \emph{{Thermodynamics of polarized relativistic matter}},
  \href{https://doi.org/10.1007/JHEP07(2016)028}{\emph{JHEP} {\bfseries 07}
  (2016) 028} [\href{https://arxiv.org/abs/1606.01226}{{\ttfamily
  1606.01226}}].

\bibitem{Hernandez:2017mch}
J.~Hernandez and P.~Kovtun, \emph{{Relativistic magnetohydrodynamics}},
  \href{https://doi.org/10.1007/JHEP05(2017)001}{\emph{JHEP} {\bfseries 05}
  (2017) 001} [\href{https://arxiv.org/abs/1703.08757}{{\ttfamily
  1703.08757}}].

\bibitem{Gedalin1993}
M.~Gedalin, \emph{Linear waves in relativistic anisotropic
  magnetohydrodynamics},
  \href{https://doi.org/10.1103/PhysRevE.47.4354}{\emph{Phys. Rev. E}
  {\bfseries 47} (1993) 4354}.

\bibitem{Bekenstein2006}
J.D.~Bekenstein and G.~Betschart, \emph{Perfect magnetohydrodynamics as a field
  theory}, \href{https://doi.org/10.1103/PhysRevD.74.083009}{\emph{Phys. Rev.
  D} {\bfseries 74} (2006) 083009}
  [\href{https://arxiv.org/abs/gr-qc/0608053}{{\ttfamily gr-qc/0608053}}].

\bibitem{Aguiar:2003pp}
C.~Aguiar, E.~Fraga and T.~Kodama, \emph{{Hydrodynamical instabilities beyond
  the chiral critical point}},
  \href{https://doi.org/10.1088/0954-3899/32/2/009}{\emph{J. Phys. G}
  {\bfseries 32} (2006) 179}
  [\href{https://arxiv.org/abs/nucl-th/0306041}{{\ttfamily nucl-th/0306041}}].

\bibitem{Friedman2013}
J.L.~Friedman and N.~Stergioulas, \emph{Rotating Relativistic Stars}, Cambridge
  Monographs on Mathematical Physics, Cambridge University Press (3, 2013),
  \href{https://doi.org/10.1017/CBO9780511977596}{10.1017/CBO9780511977596}.

\bibitem{Jensen2013}
K.~Jensen, R.~Loganayagam and A.~Yarom, \emph{Anomaly inflow and thermal
  equilibrium},  \href{https://arxiv.org/abs/1310.7024 [hep-th]}{{\ttfamily
  1310.7024 [hep-th]}}.

\bibitem{Kapusta2011}
J.~Kapusta and C.~Gale, \emph{Finite-temperature Field Theory: Principles and
  Applications}, Cambridge Monographs on Mathematical Physics, Cambridge
  University Press (2011),
  \href{https://doi.org/10.1017/CBO9780511535130}{10.1017/CBO9780511535130}.

\bibitem{Israel:1976tn}
W.~Israel, \emph{{Nonstationary irreversible thermodynamics: A Causal
  relativistic theory}},
  \href{https://doi.org/10.1016/0003-4916(76)90064-6}{\emph{Annals Phys.}
  {\bfseries 100} (1976) 310}.

\bibitem{Becattini:2012tc}
F.~Becattini, \emph{{Covariant statistical mechanics and the stress-energy
  tensor}}, \href{https://doi.org/10.1103/PhysRevLett.108.244502}{\emph{Phys.
  Rev. Lett.} {\bfseries 108} (2012) 244502}
  [\href{https://arxiv.org/abs/1201.5278}{{\ttfamily 1201.5278}}].

\bibitem{Zee2013}
A.~Zee, \emph{Einstein Gravity in a Nutshell}, Princeton University Press, New
  Jersey (5, 2013).

\bibitem{DeGroot1980}
S.~De~Groot, \emph{Relativistic Kinetic Theory. Principles and Applications}
  (1, 1980).

\bibitem{Pu:2009fj}
S.~Pu, T.~Koide and D.H.~Rischke, \emph{{Does stability of relativistic
  dissipative fluid dynamics imply causality?}},
  \href{https://doi.org/10.1103/PhysRevD.81.114039}{\emph{Phys. Rev. D}
  {\bfseries 81} (2010) 114039}
  [\href{https://arxiv.org/abs/0907.3906}{{\ttfamily 0907.3906}}].

\bibitem{Routh}
E.J.~Routh, \emph{A treatise on the stability of a given state of motion:
  Particularly steady motion}, Macmillan (1877).

\bibitem{Hurwitz1895}
A.~Hurwitz, \emph{Ueber die bedingungen, unter welchen eine gleichung nur
  wurzeln mit negativen reellen theilen besitzt},
  \href{https://doi.org/10.1007/BF01446812}{\emph{Mathematische Annalen}
  {\bfseries 46} (1895) 273}.

\bibitem{Krotscheck1978}
E.~Krotscheck and W.~Kundt, \emph{Causality criteria},
  \href{https://doi.org/10.1007/BF01609447}{\emph{Communications in
  Mathematical Physics} {\bfseries 60} (1978) 171}.

\bibitem{Pesic2004}
P.~{Pesic}, \emph{{Abel's proof. An essay on the sources and meaning of
  mathematical unsolvability. Reprint}}, Cambridge, MA: MIT Press (2004).

\bibitem{Kawazura:2017lpc}
Y.~Kawazura, \emph{{Modification of magnetohydrodynamic waves by the
  relativistic Hall effect}},
  \href{https://doi.org/10.1103/PhysRevE.96.013207}{\emph{Phys. Rev. E}
  {\bfseries 96} (2017) 013207}
  [\href{https://arxiv.org/abs/1706.07077}{{\ttfamily 1706.07077}}].

\bibitem{Liu2016}
C.-X.~Liu, S.-C.~Zhang and X.-L.~Qi, \emph{The quantum anomalous hall effect:
  Theory and experiment},
  \href{https://doi.org/10.1146/annurev-conmatphys-031115-011417}{\emph{Annual
  Review of Condensed Matter Physics} {\bfseries 7} (2016) 301}
  [\href{https://arxiv.org/abs/https://doi.org/10.1146/annurev-conmatphys-031115-011417}{{\ttfamily
  https://doi.org/10.1146/annurev-conmatphys-031115-011417}}].

\end{thebibliography}\endgroup
\end{document}